\documentclass[twoside,twocolumn,9pt]{article}
\usepackage{extsizes}
\usepackage[super,sort&compress,comma]{natbib} 
\usepackage[version=3]{mhchem}
\usepackage[left=1.5cm, right=1.5cm, top=1.785cm, bottom=2.0cm]{geometry}
\usepackage[mathlines]{lineno}% Enable numbering of text and display math
\usepackage{times,mathptmx}
\usepackage{sectsty}
\usepackage[pdftex]{graphicx} 
\usepackage[format=plain,justification=raggedright,singlelinecheck=false,font={stretch=1.125,small,sf},labelfont=bf,labelsep=space]{caption}
\usepackage{float}
\usepackage{fancyhdr}
\usepackage{fnpos}
\usepackage[english]{babel}
\usepackage{array}
\usepackage{droidsans}
\usepackage{charter}
\usepackage[T1]{fontenc}
\usepackage[usenames,dvipsnames]{xcolor}
\usepackage{setspace}
\usepackage[compact]{titlesec}
\usepackage{epstopdf}%This line makes .eps figures into .pdf - please comment out if not required.

\definecolor{cream}{RGB}{222,217,201}

\newcommand{\hh}{H$_4$ }

\newcommand{\be}{\begin{equation}}
\newcommand{\ee}{\end{equation}}
\newcommand{\bea}{\begin{eqnarray}}
\newcommand{\eea}{\end{eqnarray}}

\newcommand{\ie}{{\it i.e.,\ }}

\newcommand{\half}{\frac12}

\newcommand{\rr}{\mathbf{1}}
\newcommand{\rt}{\mathbf{2}}

\newcommand{\ND}{\text{ND}}
\newcommand{\D}{\text{D}}
\newcommand{\T}{\text{T}}
\newcommand{\HH}{\text{H}}
\newcommand{\CAS}{\text{CASSCF}}
\newcommand{\FCI}{\text{FCI}}
\newcommand{\HF}{\text{HF}}
\newcommand{\HFL}{\text{HFL}}
\newcommand{\PHF}{\text{PHF}}

\newcommand{\CORR}{\text{CORR}}
\newcommand{\DD}{\mathbf{D}}
\newcommand{\s}{\sigma}

\begin{document}

\pagestyle{fancy}
\thispagestyle{plain}
\fancypagestyle{plain}{

%%%HEADER%%%
\fancyhead[C]{\includegraphics[width=18.5cm]{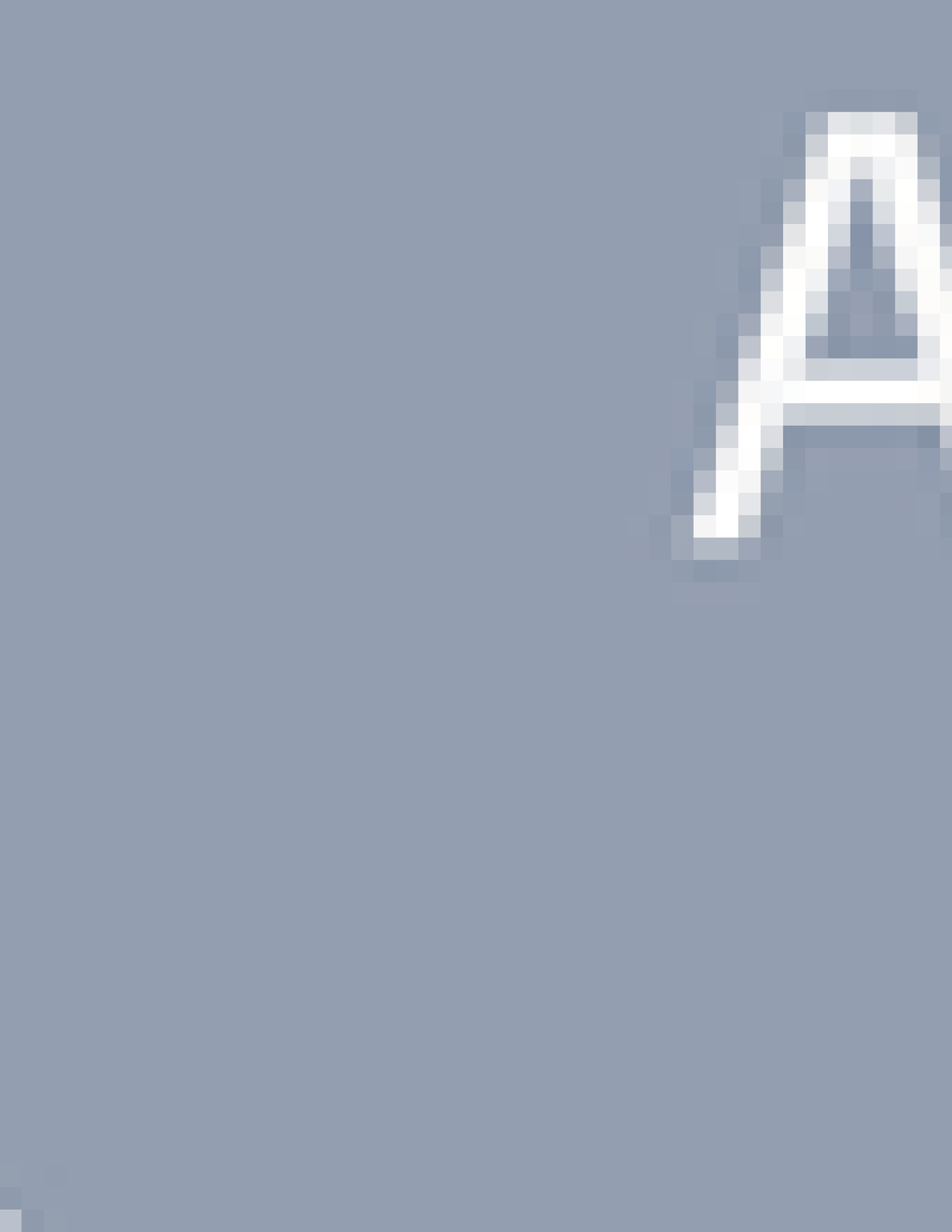}}
\fancyhead[L]{\hspace{0cm}\vspace{1.5cm}\includegraphics[height=30pt]{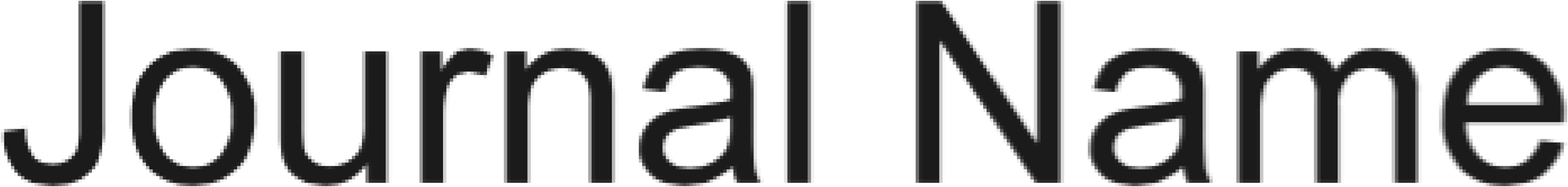}}
\fancyhead[R]{\hspace{0cm}\vspace{1.7cm}\includegraphics[height=55pt]{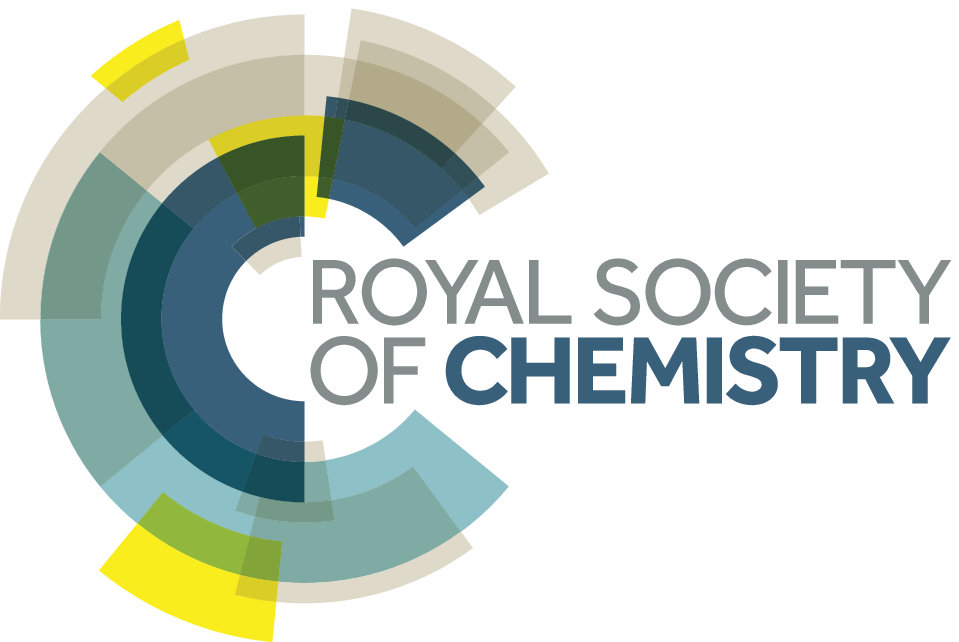}}
\renewcommand{\headrulewidth}{0pt}
}
%%%END OF HEADER%%%

%%%PAGE SETUP - Please do not change any commands within this section%%%
\makeFNbottom
\makeatletter
\renewcommand\LARGE{\@setfontsize\LARGE{15pt}{17}}
\renewcommand\Large{\@setfontsize\Large{12pt}{14}}
\renewcommand\large{\@setfontsize\large{10pt}{12}}
\renewcommand\footnotesize{\@setfontsize\footnotesize{7pt}{10}}
\makeatother

\renewcommand{\thefootnote}{\fnsymbol{footnote}}
\renewcommand\footnoterule{\vspace*{1pt}% 
\color{cream}\hrule width 3.5in height 0.4pt \color{black}\vspace*{5pt}} 
\setcounter{secnumdepth}{5}

\makeatletter 
\renewcommand\@biblabel[1]{#1}            
\renewcommand\@makefntext[1]% 
{\noindent\makebox[0pt][r]{\@thefnmark\,}#1}
\makeatother 
\renewcommand{\figurename}{\small{Fig.}~}
\sectionfont{\sffamily\Large}
\subsectionfont{\normalsize}
\subsubsectionfont{\bf}
\setstretch{1.125} %In particular, please do not alter this line.
\setlength{\skip\footins}{0.8cm}
\setlength{\footnotesep}{0.25cm}
\setlength{\jot}{10pt}
\titlespacing*{\section}{0pt}{4pt}{4pt}
\titlespacing*{\subsection}{0pt}{15pt}{1pt}
%%%END OF PAGE SETUP%%%

%%%FOOTER%%%
\fancyfoot{}
\fancyfoot[LO,RE]{\vspace{-7.1pt}\includegraphics[height=9pt]{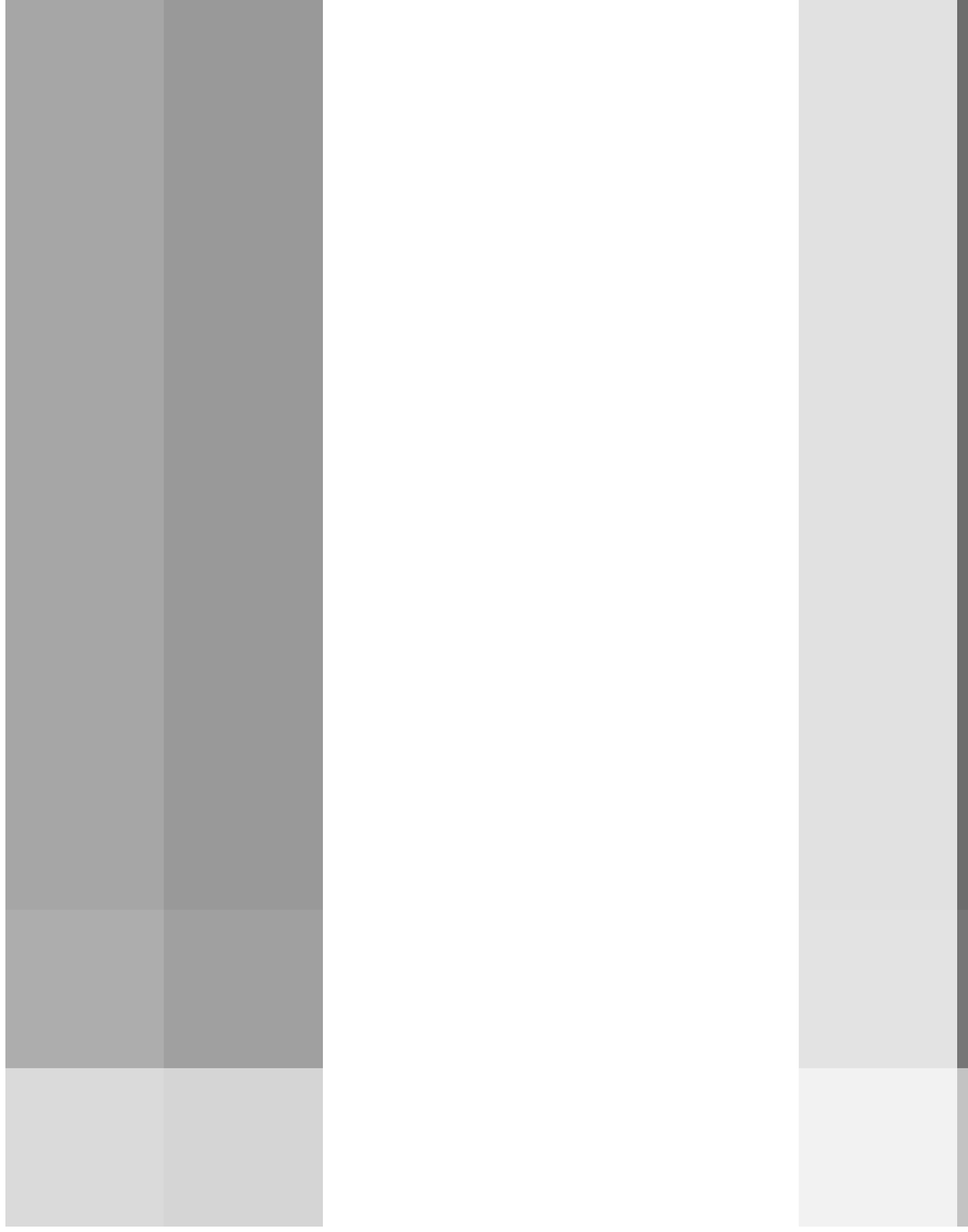}}
\fancyfoot[CO]{\vspace{-7.1pt}\hspace{13.2cm}\includegraphics{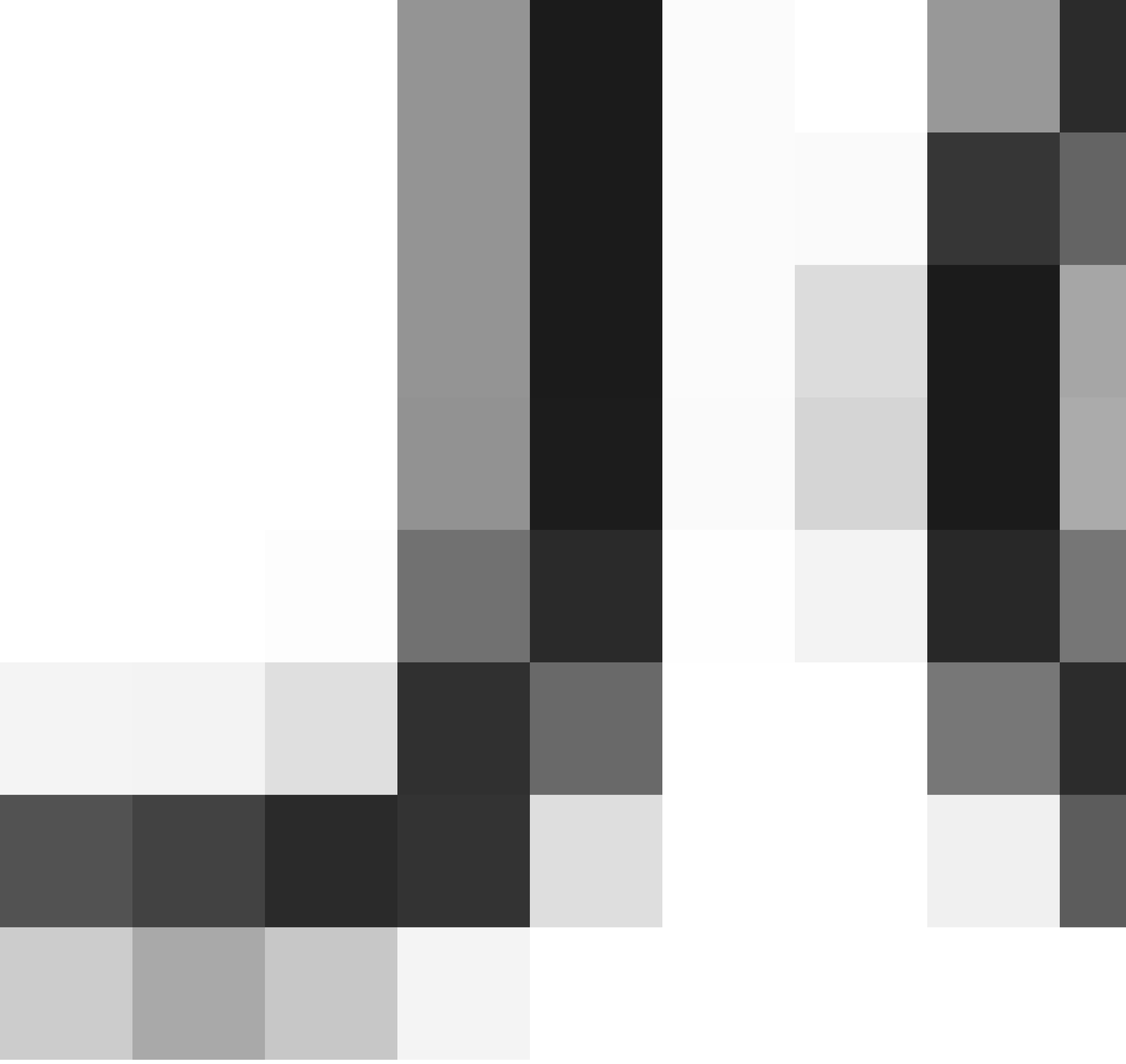}}
\fancyfoot[CE]{\vspace{-7.2pt}\hspace{-14.2cm}\includegraphics{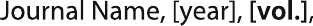}}
\fancyfoot[RO]{\footnotesize{\sffamily{1--\pageref{LastPage} ~\textbar  \hspace{2pt}\thepage}}}
\fancyfoot[LE]{\footnotesize{\sffamily{\thepage~\textbar\hspace{3.45cm} 1--\pageref{LastPage}}}}
\fancyhead{}
\renewcommand{\headrulewidth}{0pt} 
\renewcommand{\footrulewidth}{0pt}
\setlength{\arrayrulewidth}{1pt}
\setlength{\columnsep}{6.5mm}
\setlength\bibsep{1pt}
%%%END OF FOOTER%%%

%%%FIGURE SETUP - please do not change any commands within this section%%%
\makeatletter 
\newlength{\figrulesep} 
\setlength{\figrulesep}{0.5\textfloatsep} 

\newcommand{\topfigrule}{\vspace*{-1pt}% 
\noindent{\color{cream}\rule[-\figrulesep]{\columnwidth}{1.5pt}} }

\newcommand{\botfigrule}{\vspace*{-2pt}% 
\noindent{\color{cream}\rule[\figrulesep]{\columnwidth}{1.5pt}} }

\newcommand{\dblfigrule}{\vspace*{-1pt}% 
\noindent{\color{cream}\rule[-\figrulesep]{\textwidth}{1.5pt}} }

\makeatother
%%%END OF FIGURE SETUP%%%

%%%TITLE, AUTHORS AND ABSTRACT%%%
\twocolumn[
  \begin{@twocolumnfalse}
\vspace{3cm}
\sffamily
\begin{tabular}{m{4.5cm} p{13.5cm} }

\includegraphics{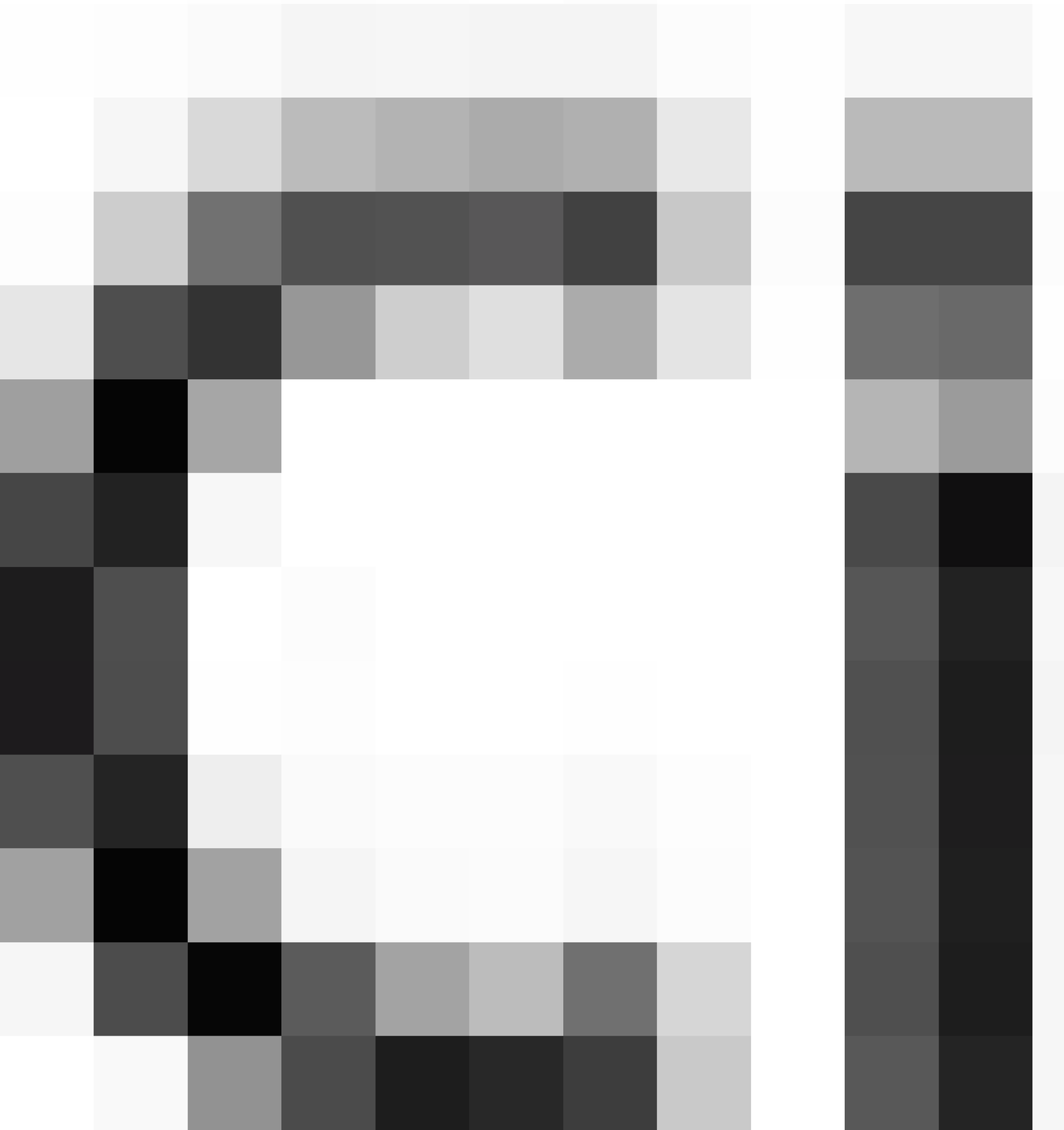} & \noindent\LARGE{\textbf{Separation of Dynamic and Nondynamic Correlation}} \\
\vspace{0.3cm} & \vspace{0.3cm} \\

 & \noindent\large{\textbf{Eloy Ramos-Cordoba$^{\ast}$\textit{$^{a,b}$}, Pedro Salvador\textit{$^{c}$} and Eduard Matito$^{\ast}$\textit{$^{a,d}$}}}\vspace{0.5cm} \\

\includegraphics{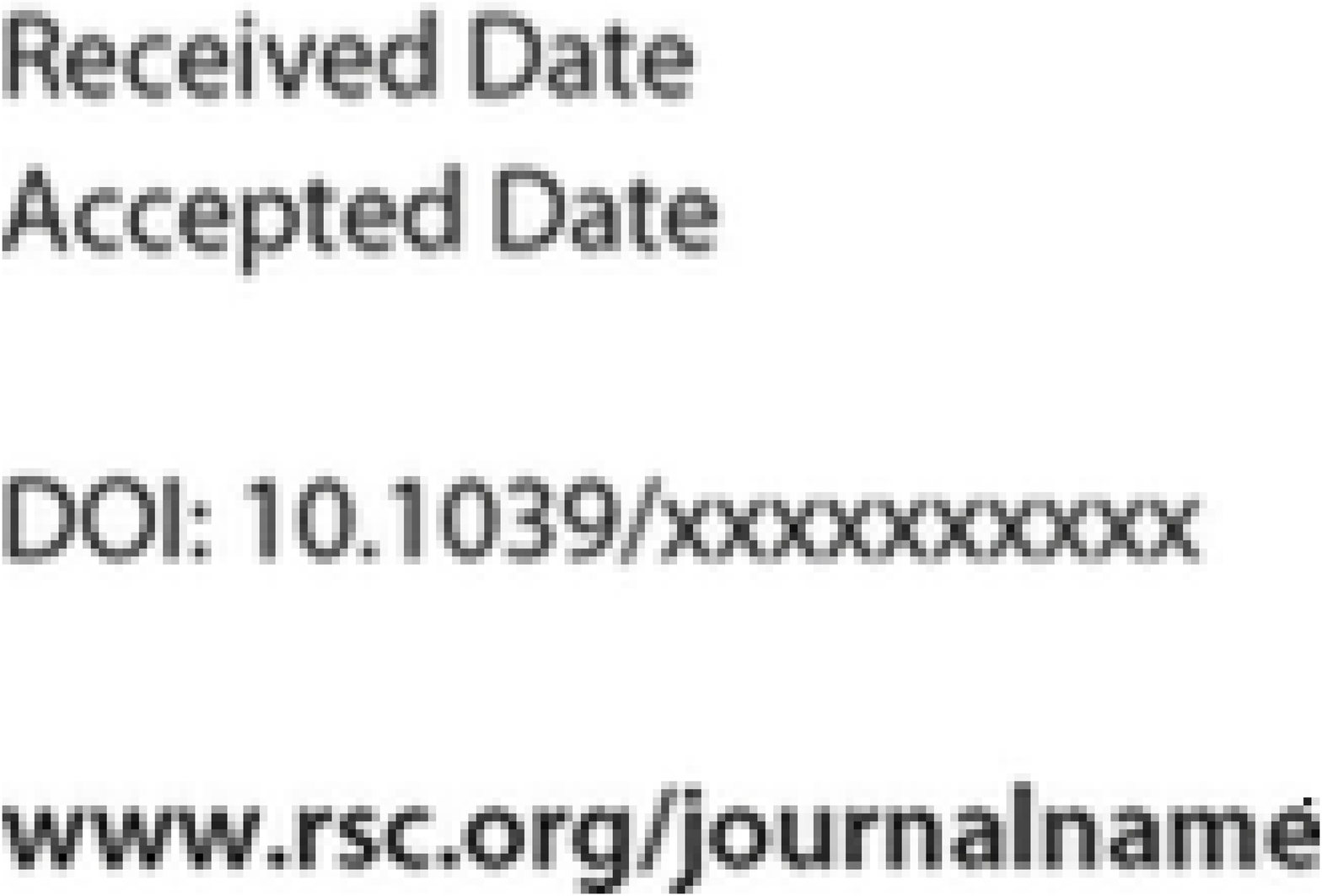} & \noindent\normalsize{
The account of electron correlation and its efficient 
separation into dynamic and nondynamic parts plays a key role in the development
of computational methods. 
In this paper we suggest a physically-sound matrix formulation to split electron correlation 
into dynamic and nondynamic parts using
the two-particle cumulant matrix and 
a measure of the deviation from idempotency of the first-order density matrix.
These matrices are applied to a
two-electron model, giving rise to a simplified electron correlation index that 
($i$) depends only on natural
orbitals and their occupancies, ($ii$) can be straightforwardly decomposed into orbital contributions
and ($iii$) splits into dynamic and nondynamic correlation parts
that ($iv$) admit a local version.
These expressions are shown to account for dynamic and nondynamic correlation
in a variety of systems containing different electron correlation regimes, thus
providing the first separation of dynamic and nondynamic correlation using solely natural
orbital occupancies.
} \\

\end{tabular}

 \end{@twocolumnfalse} \vspace{0.6cm}

  ]
%%%END OF TITLE, AUTHORS AND ABSTRACT%%%

%%%FONT SETUP - please do not change any commands within this section
\renewcommand*\rmdefault{bch}\normalfont\upshape
\rmfamily
\section*{}
\vspace{-1cm}

%%%FOOTNOTES%%%

\footnotetext{\textit{$^{a}$~Faculty of Chemistry, University of the Basque Country UPV/EHU,
and Donostia International Physics Center (DIPC). 
P.K. 1072, 20080 Donostia, Euskadi, Spain.}}
\footnotetext{\textit{$^{b}$~Kenneth S. Pitzer Center for Theoretical Chemistry, UC Berkeley, USA}}
\footnotetext{\textit{$^{c}$~Institut de Qu\'imica Computacional i Cat{\`a}lisi (IQCC) and
Departament de Qu\'imica, University of Girona, 17071 Girona, Catalonia, Spain.}}
\footnotetext{\textit{$^{d}$~IKERBASQUE, Basque Foundation for Science, 48011 Bilbao, Spain.}}

%Please use \dag to cite the ESI in the main text of the article.
%If you article does not have ESI please remove the the \dag symbol from the title and the footnotetext below.
%\footnotetext{\dag~Electronic Supplementary Information (ESI) available: [details of any supplementary information available should be included here]. See DOI: 10.1039/b000000x/}
%additional addresses can be cited as above using the lower-case letters, c, d, e... If all authors are from the same address, no letter is required

%\footnotetext{\ddag~Additional footnotes to the title and authors can be included \emph{e.g.}\ `Present address:' or `These authors contributed equally to this work' as above using the symbols: \ddag, \textsection, and \P. Please place the appropriate symbol next to the author's name and include a \texttt{\textbackslash footnotetext} entry in the the correct place in the list.}

%%%END OF FOOTNOTES%%%

%%%MAIN TEXT%%%%

%
%\DeclareMathOperator{\Tr}{Tr}

\section{Introduction}
The account of electron correlation effects ---\textit{i.e.}, the interaction between electrons
in a quantum system--- is still a most important challenge in current computational 
chemistry.~\cite{hattig:12cr,cremer:01mp,ziesche:00the,tew:07jcc} 
Many different properties are affected by electron correlation, including 
bond stretching and dissociation, electron delocalization or antiferromagnetic interactions.
The issue has been addressed from multiple perspectives, giving rise to a manifold of methods to calculate the energy of a molecular system. Each method has its own way to tackle the problem and, to some extent, recover electron correlation effects. 
There are multiple phenomena behind electron correlation, which have been largely studied in the past sixty 
years.~\cite{lowdin:55pr,cremer:01mp,ziesche:00the,tew:07jcc,kutzelnigg:68pr,raghavachari:96jpc,juhasz:06jcp,alcoba:10jcp,skolnik:13pra}
As a result, nowadays terms such as short-range, long-range, 
dynamic,~\cite{sinanoglu:64acp} nondynamic, static,~\cite{bartlett:07rcc}
left-right, in-out, radial or angular~\cite{lennardjones:52jcp} 
correlation belong to the jargon that computational
chemists use to analyze the missing electron correlation effects of a given method in the 
description of quantum systems.\newline

Perhaps the most well-known classification of electron correlation types is done in terms of \text{dynamic} 
and \text{nondynamic} electron correlation.~\cite{sinanoglu:64acp}
A method is said to include \textit{dynamic correlation} if its wavefunction, 
calculated as a configuration interaction (CI) expansion, includes a most dominant configuration and only small 
(but energetically important) contributions of other configurations. 
Conversely, \textit{nondynamic correlation} arises when there are two or more important configurations, 
usually in the presence of degeneracies or near-degeneracies. 
Nondynamic correlation is regarded as a system-specific contribution, whereas dynamic electron
correlation is accepted to be a rather \textit{universal} contribution.~\cite{lie:74jcp,sinanoglu:64acp}
Some authors~\cite{bartlett:07rcc} distinguish between nondynamic and static
correlation and define \textit{static correlation} as the nondynamic correlation
required to provide a correct zeroth-order description as dictated by spin
and symmetry considerations, whereas nondynamic correlation is reserved for
other situations such as the separation of molecules into fragments or the
description of some excited states. In this paper we will use the term nondynamic
correlation in a wide sense, comprising both static and (pure) nondynamic effects,
\ie all the electron correlation that is not dynamic.
Interestingly, many popular methods introduce 
mainly either one or the other correlation type, and 
therefore, a large number of computational approaches can be classified as either dynamic- or nondynamic-correlation including methods. 
For instance, truncated CI and coupled-cluster (CC) wavefunctions give a good account of dynamic correlation, 
while multiconfiguration (MC) wavefunctions introduce mostly nondynamic correlation. 
%Most computational methods are 
%efficient at introducing one or the other electron correlation type, but not both simultaneously. 
Back in the 90's different diagnostic tools were put forward in order to
evaluate the extent of dynamic and nondynamic correlation included in
computational methods.
For instance,
the so-called T1 diagnostic of Lee and coworkers,~\cite{lee:89tca,lee:89ijqc} 
the less-known D1 diagnostic~\cite{janssen:98cpl} and subsequent 
modifications~\cite{lee:03cpl}
have been widely used to measure the importance of nondynamic 
correlation in coupled-cluster wave functions. 
Cioslowski also developed the differential density matrix overlap (DDMO)
for the assessment of electron correlation effects.~\cite{cioslowski:92tca}
The development of new cost-efficient computational methods has often consisted in the inclusion of 
dynamic or nondynamic correlation to an existing approach that already incorporates the other correlation 
type. This has given rise to a plethora of procedures such as second-order perturbation theory from
complete active space wavefunctions (CASPT2),~\cite{andersson:92jcp} multireference configuration
interactions with single and double excitations (MRCI-SD)~\cite{buenker:78mp}
The mixing of two existing computational approximations have also resulted in the construction of 
\textit{hybrid methods}, 
often combining density functional theory (DFT) and wavefunction techniques.~\cite{savin:88ijqc,savin:91chapter}
Range-separated methods,~\cite{savin:96bc} which provide a scheme to merge short and long-range
correlation, or (local)~\cite{jaramillo:03jcp} hybrid density functionals,~\cite{becke:93jcp,becke:93jcp1} 
which use orbital-based Hartree-Fock exchange
mixed with DFT exchange 
are just two examples. These methods
use optimally-chosen mixing parameters or local mixing functions (LMF) to ponderate
the ingredients combined. The mixture of exact exchange and density functional
approximations should be determined according to the properties of 
each system~\cite{perdew:96jcp} or, even better, using a LMF that depends
on the features of the target molecule.~\cite{cruz:98jpca,jaramillo:03jcp}
The admixture is rationalized as a balance between the simulation of long-range
nondynamic correlation effects (included by the DFT exchange, usually through the 
generalized-gradient approximation) and the self-interaction correction 
(exact exchange),~\cite{cremer:01mp} whereas the inclusion of dynamic correlation effects
is given by a local correlation functional multiplied by a mixing parameter.
Range-separated functionals~\cite{iikura:01jcp} depend on attenuating parameters that
have been likewise shown to be system specific.~\cite{baer:09arpc}
The key question in local hybrid functionals is the choice of the LMF, determining the
local relative contributions of exact and DFT exchange. 
Several possibilities have been explored,~\cite{arbuznikov:14jcp,henderson:08jpca} 
including the ratio of von Weizs\"acker and the exact kinetic
energy densites~\cite{jaramillo:03jcp} and the correlation length,~\cite{johnson:14jcp}
which are related
to the \textit{local} contribution to electron correlation.~\cite{becke:88jcp}
Despite the promising results obtained, no optimal LMF providing good results
in both thermodynamics and kinetic benchmark tests has been found thus 
far.~\cite{johnson:14jcp,silva:15jcp}\newline

The account of electron correlation and its efficient 
separation into dynamic and nondynamic effects thus plays a key role in many situations.
Grimme et al. have recently put forward a local measure of nondynamic correlation based on fractional
orbital occupations within finite-temperature DFT,~\cite{grimme:15ang} Reiher and coworkers suggested
orbital entaglement measures to evaluate the nondynamic correlation in density-matrix
renormalization group (DMRG)~\cite{boguslawski:12jpcl} and
Raeber and Mazziotti have proposed
a global indicator of the off-diagonal long-range order correlation.~\cite{raeber:15pra}
To our knowledge, it does not exist a mesure of electron correlation that can be split into
dynamic and nondynamic counterparts and admits a local decomposition. Such indicator
can be obviously used to assess the relative importance of dynamic and nondynamic
correlation effects but it can also aid the development of new computational methods
either by exploting the local character of dynamic electron correlation in local
methods~\cite{pulay:83cpl,zalesny:11book}
or by affording a new means to mix the components of hybrid methods.~\cite{savin:96bc}
In this paper we develop a general expression to account for electron correlation effects
that can be decomposed into dynamic and nondynamic parts. The expression is applied to a
two-electron model, giving rise to a simplified expression that ($i$) depends only on natural
orbitals and their occupancies, ($ii$) can be straightforwardly decomposed into orbital contributions,
($iii$) can be split into dynamic and nondynamic correlation contributions
and ($iv$) all its contributions admit a local version.
This expression is finally shown to account for dynamic and nondynamic correlation in several
molecular systems, thus validating its applicability beyond the model system.

\section{Precedents}
\begin{figure}[h]
\begin{center}
\includegraphics[width=0.28\textwidth]{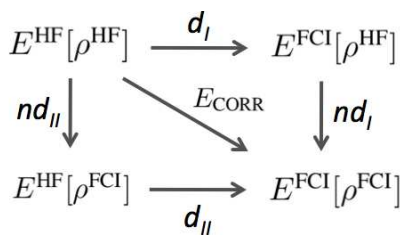}
\end{center}
\caption{Correlation energy diagram. 
Path I follows Cioslowski's decomposition of the electron correlation and
path II is due to Lude{\~n}a and coworkers.
Adapted from Ref.~\citenum{valderrama:97jcp}.}
\label{f:EcorrDiag}
\end{figure}

It is not straightforward, and perhaps even impossible, to make a clear-cut separation between dynamic 
and nondynamic electron correlation.~\cite{shavitt:84chapter}
The increase of electron excitations in a CI or CC 
wavefunction eventually introduces some nondynamic correlation effects, whereas the increase of 
the number of configurations in a MC wavefunction at some point should also include dynamic correlation. 
However, since the source of these electron correlation types 
 (and, therefore the way to account for them) 
is so different, it becomes essential to have simple expressions to distinguish one from the other.
Unfortunately,
there are few simple computational expressions in the literature 
that can provide a quantitative analysis of dynamic and nondynamic correlation effects. 
One of the first such separations 
is due to Cioslowski~\cite{cioslowski:91pra} and it was, subsequently, generalized by 
Lude{\~n}a and coworkers.~\cite{valderrama:97jcp,valderrama:99jcp}
Fig.~\ref{f:EcorrDiag} summarizes the main idea.
The uppermost left corner corresponds to the Hartree-Fock (HF) energy, which is connected to the opposite corner 
---the non-relativistic full-CI (FCI) energy--- by the correlation energy. 
One can also cover the distance between HF and FCI in two steps by following two different paths, 
\begin{eqnarray}\label{eq:ludena}
E_{\CORR} = E_{\ND}^{(\text{I})}+E_{D}^{(\text{I})}=E_{D}^{(\text{II})}+E_{\ND}^{(\text{II})} \,\, .
\end{eqnarray}
Cioslowski~\cite{cioslowski:91pra} suggested to calculate the FCI energy with a density-constrained approach that selects the
FCI wavefunction reproducing the HF density, \textit{i.e.}, $E^{\FCI}[\rho^{\HF}]$. 
The step from $E^{\HF}$ to $E^{\FCI}[\rho^{\HF}]$ corresponds to path $d_\text{I}$ 
in Fig.~\ref{f:EcorrDiag}, and it is expected to retrieve only the dynamic correlation
because it uses the expression for 
an exact wavefunction but its electron density is restricted to the HF one. On the contrary, 
the relaxation process that permits global changes in the 2-DM as the electron
density transforms 
from HF density to FCI one (path $nd_\text{I}$), should account for nondynamic correlation. 
Lude{\~n}a suggested 
the path II that goes through 
$E^{\HF}[\rho^{\FCI}]$, which is a HF calculation using a single-determinant wavefunction
restricted to reproduce the FCI density. Path II recovers first 
nondynamic and afterwards dynamic correlation.
Cioslowski's definition is preferred because it provides a decomposition of electron
correlation into nonpositive contributions, whereas Lude{\~n}a's definition provides
nonnegative electron correlation values, namely, $nd_{\text{II}}>0$.
Notwithstanding, both approaches afford an exact
decomposition of the electron correlation energy and provide a similar description of 
dynamic and nondynamic correlation.~\cite{valderrama:97jcp,valderrama:99jcp}
Despite the physically-soundness of these methods, in practice, they require
expensive FCI calculations that prevent their application beyond few-electron
species.
A more practical realization of dynamic/nondynamic separation is due to
Handy and coworkers,~\cite{mok:96jpc} who suggested that a
full-valence complete active space self-consistent field (CASSCF) calculation could 
be used as a reference wavefunction that only includes the nondynamic correlation 
energy, giving rise to the following separation of dynamic and nondynamic correlation energy,\newline
\begin{eqnarray}\nonumber
E_{\CORR} &=& E_{\FCI}-E_{\HF}=E_{\ND}^{\HH}+E_{\D}^{\HH} \\
&=& \left(E_{\CAS}-E_{\HF}\right)+\left(E_{\FCI}-E_{\CAS}\right)
\label{eq:handy}
\end{eqnarray}
This expression for dynamic correlation energy actually coincides with the electron
correlation energy definition suggested by Davidson some years earlier.~\cite{davidson:91pra}\newline

Cioslowski's, Handy's and Lude{\~na}'s are schemes that
afford the separation of dynamic and nondynamic correlation energies.
There are very few other quantitative
indicators of electron correlation and, to the best of our knowledge, there is no other 
measure that can quantify the relative importance of both effects. 
Ugalde and coworkers suggested the use of the intracule densities to obtain
a separation of dynamic and nondynamic correlation effects~\cite{valderrama:01jpb}
but their approach, in the spirit of Handy's, also relies on a user-defined 
wavefunction that contains nondynamic electron correlation effects.\newline

It would be more convenient if the separation of dynamic and
nondynamic correlation would not depend on the definition of a wavefunction
containing only nondynamic correlation effects.
First of all, such definition is always arbitrary and,
in some cases (see the discussion on H$_2$, LiH and Li$_2$ below), 
difficult to construct.
Besides, an approach that allows the separation
of electron correlation rather than an energy-decomposition scheme, could be used
to split the correlation contributions
of other observables.
Ideally, 
the measure should depend on very simple quantities. For instance, 
expressions based on natural orbital occupancies (NOO) would accomplish this goal and 
would be directly applicable to natural orbital and density matrix functional 
theories~\cite{piris:14ijqc,pernal:15tcc}
 as well as wavefunction ab initio calculations.~\cite{hattig:12cr}
In addition, through appropriate transformations, the NOO-based
measures of electron correlation could also be
applied in density functional theory.~\cite{gruning:03jcp,wu:15jctc} One such measure is
the deviation
from idempotency (DFI)~\cite{lowdin:55pr,smith:67tca} that uses NOO
and accounts for nondynamic correlation effects. 
There is actually
very few indicators of dynamic electron correlation available in the 
literature and, to our knowledge, 
there is no NOO-based dynamic correlation measure excepting for Ziesche's 
proposal,~\cite{ziesche:00the}
which uses NOO of orbitals close to the Fermi level but it is not continuous
with respect to small changes of NOO.\newline

%The goal of this work is introducing a new dynamic correlation index that
%will complement the DFI, providing a NOO-based separation of dynamic and
%nondynamic correlation effects.

\section{Dynamic and Nondynamic separation}
In order to construct a physically-motivated separation of dynamic and nondynamic correlation 
effects we will use the second- and first-order reduced density matrices (2-RDM and 1-RDM) 
in a natural orbital representation, respectively:
\begin{eqnarray}
\rho_1(\rr;\rr') &=& \sum_{i} n_{i} \phi^*_i(\rr') \phi_i(\rr)
\end{eqnarray}
\begin{eqnarray}
\rho_2(\rr,\rt;\rr',\rt')&=& \sum_{ijkl} {^2D}^{ij}_{kl}\phi^*_i(\rr') \phi_j^*(\rt') \phi_k(\rr) \phi_l(\rt)
\end{eqnarray}
where $\phi_i(\rr)$ is a natural orbital and $n_i$ its occupancy, $\rr\equiv(\vec{r}_1,\s_1)$,
and 
${^2D}^{ij}_{kl}$ is the matrix representation of the 2-RDM (2-DM hereafter)
and it should contain
the electron correlation information of the system.~\cite{coulson:60rmp}
The Hartree-Fock-like approximation
of the 2-DM becomes a very simple function in terms of NOO,~\cite{lowdin:55pr}
\begin{equation}\label{eq:sd2rdm}
({^2D}^{\HFL})^{ij}_{kl}=
n_in_j\left(\delta_{ik}\delta_{jl}-\delta_{il}\delta_{jk}\right)
\end{equation}
the expression being exact for single-determinant wavefunctions.
Let us define a pseudo Hartree-Fock (HF) 2-DM in the natural orbital representation
\begin{equation}\label{eq:hf2rdm}
({^2D}^{\PHF})^{ij}_{kl}=
\xi_{ijkl}\left(\delta_{ik}\delta_{jl}-\delta_{il}\delta_{jk}\right) \, ,
\end{equation}
where $\xi_{ijkl}$ equals 1 if $i, j, k$ and $l$ orbitals are below the Fermi
level, zero otherwise. 
The difference with respect to the actual HF 2-DM is that
the pseudo-HF is defined in terms of the natural orbitals of the correlated calculation
whereas the actual HF 2-DM is defined in terms of the canonical HF orbitals
(\ie the
expression looks like Eq.~\ref{eq:hf2rdm} but $i,j,etc.$ refer to 
canonical HF orbitals).
We have checked that energy-wise the difference between both
matrices is fairly small and, for practical reasons, one can take the latter
as a matrix free of electron correlation effects. In this sense, the difference
between the actual 2-DM and the pseudo-HF one is a matrix that contains electron
correlation effects, 
\begin{equation}
C^{ij}_{kl}={^2D}^{ij}_{kl}-({^2D}^{\PHF})^{ij}_{kl} \, .
\end{equation}
In order to extract the electron correlation information from these matrices let
us decompose $\mathbf{C}$ as
\begin{equation}\label{eq:decomp}
%C^{ij}_{kl}=\Lambda^{ij}_{kl}+\Gamma^{ij}_{kl}
\mathbf{C}=\boldsymbol{\Lambda}+\boldsymbol{\Gamma}  \,\, ,
\end{equation}
where,
\begin{equation}\label{eq:lambda}
\Lambda^{ij}_{kl}=\left(n_i n_j-\xi_{ijkl}\right)\left(\delta_{ik}\delta_{jl}-
\delta_{il}\delta_{jk}\right)
\end{equation}
\begin{equation}\label{eq:gamma}
\Gamma^{ij}_{kl}={^2D}^{ij}_{kl}-\left(n_i n_j\right)\left(\delta_{ik}\delta_{jl}-
\delta_{il}\delta_{jk}\right)
\end{equation}
 $\boldsymbol{\Lambda}$ is an antisymmetric diagonal matrix that measures the pairwise
deviation of NOO from a single-determinant picture and $\boldsymbol{\Gamma}$ is the
cumulant matrix.~\cite{mazziotti:98cpl,kutzelnigg:99jcp}
\begin{figure}[h]
\vspace{3mm}
\begin{center}
\includegraphics[width=0.26\textwidth]{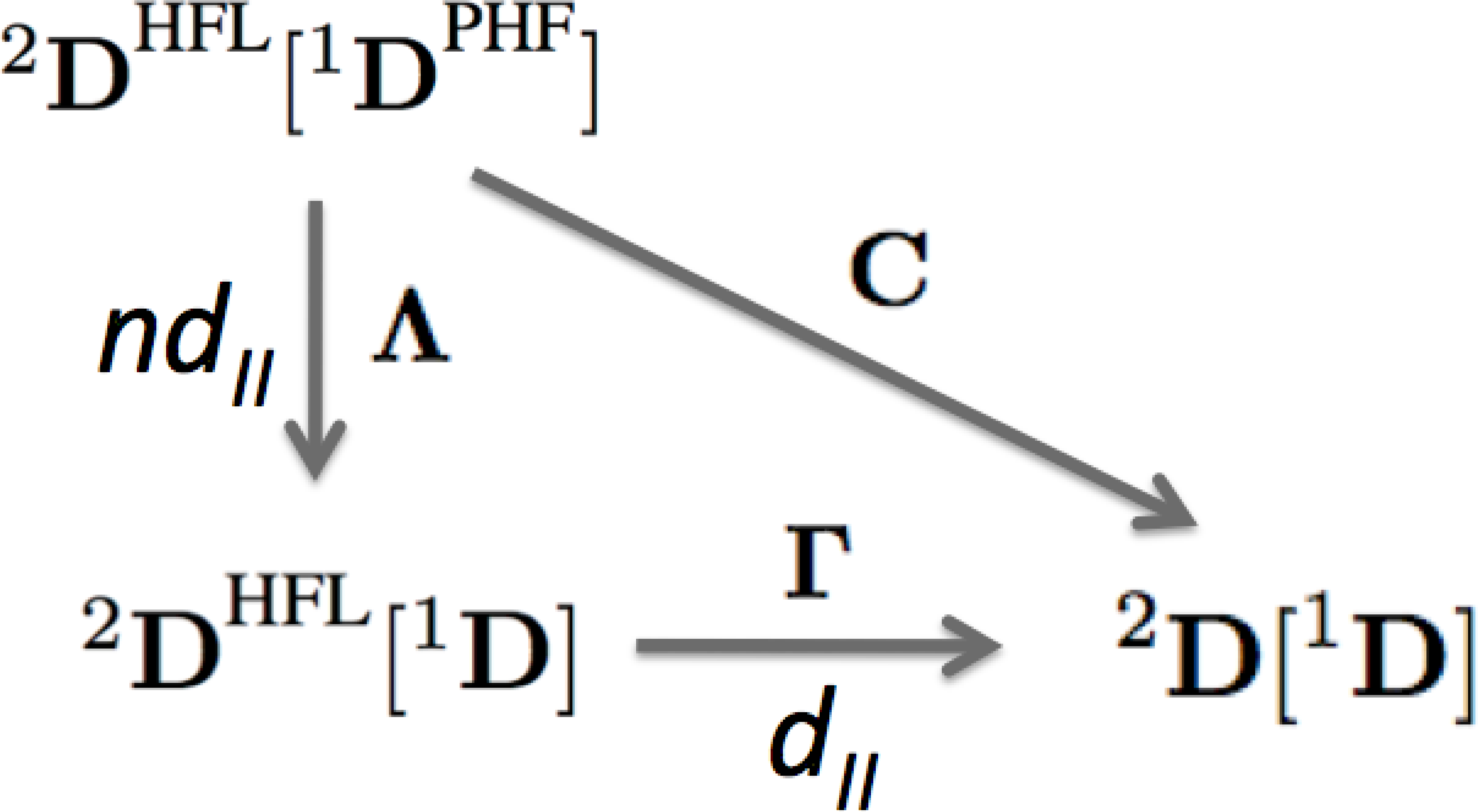}
\end{center}
\caption{Electron correlation diagram for 2-DM in terms of 1-DM.}
\label{f:DM2Diag}
\end{figure}
The decomposition of $\mathbf{C}$ is sketched in the diagram of
Fig.~\ref{f:DM2Diag}, corresponding to the two steps from the pseudo-HF 2-DM to 
the actual 2-DM.
In this diagram ${^1\DD}^{\PHF}$ is the pseudo-HF 1-DM
that we use to obtain ${^2\DD}^{\PHF}$ in Eq.~\ref{eq:hf2rdm}. The 
arrow pointing downwards represents the change in the 2-DM approximation
due to the use of \textit{exact} NOOs instead of the pseudo-HF ones. This
relaxation process is expected to retrieve nondynamic correlation effects and
is accounted by $\boldsymbol{\Lambda}$, \ie 
${^2\DD}^{\PHF}={^2\DD}^{\HFL}\left[{^1\DD}^{\PHF}\right]$.
The arrow that points from ${^2\DD}^{\HFL}[{^1\DD}]$, Eq.~\ref{eq:sd2rdm},
to the exact 2-DM represents the dynamic correlation
effects contained in $\boldsymbol{\Gamma}$. Notice that the exact electron 
density is retrieved by one-coordinate reduction of both ${^2\DD}^{\HFL}[{^1\DD}]$ 
and ${^2\DD}[{^1\DD}]$ but it is not obtained from ${^2\DD}^{\HFL}[{^1\DD^{\PHF}}]$.
Since nondynamic correlation is expected to produce global changes to the electron
density and dynamic correlation rather small local ones, it is only natural
to assign $\boldsymbol{\Lambda}$ and $\boldsymbol{\Gamma}$ as the matrices accounting for
nondynamic and dynamic correlation effects, respectively.
\newline

Eq.~\ref{eq:decomp} provides a decomposition of a matrix that contains electron
correlation into dynamic and nondynamic parts. How do we extract the information
about electron correlation from these matrices? To this aim, in the next section
we will use a two-electron model system.

\section{Two-electron model}
\begin{figure}[h]
\begin{center}
\includegraphics[width=0.32\textwidth]{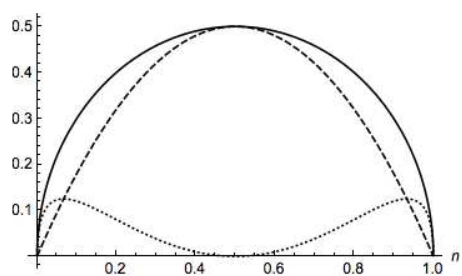}
\end{center}
\caption{Total (solid), dynamic (dotted) and nondynamic (dashed) 
electron correlation indicators for a homonuclear minimal-basis singlet two-electron model
against the orbital occupation.}
\label{f:2eMB}
\end{figure}

Let us analyze the correlation matrices defined in the latter section on a simple
model system (MS) consisting of a singlet two-electron system.
We will use a minimal basis (two orbitals) because the sign of the natural
orbitals amplitude is completely determined in this case~\cite{pernal:04jcp}
and, therefore, the 2-RDM that can be explicitly
written in terms of natural occupancies.~\cite{lowdin:56pr}
Let us consider a separation of the physical space into two symmetric regions
$F$ and $F'$, each containing an average of one electron. The electron fluctuation
between these regions can be measured through the 
covariance:~\cite{bader:74cpl,matito:07fd}
\begin{equation}
V_{F,F'}=
\left\langle \left(\widehat{N}_F-\overline{N}_F\right)\left(\widehat{N}_{F'}-\overline{N}_{F'}\right)\right\rangle
= \left\langle \widehat{N}_F\widehat{N}_{F'}\right\rangle-1
\end{equation}
where $\widehat{N}_F$ is the particle number operator acting on region $F$, 
$\overline{N}_F$ is average number of electrons in $F$ and 
$\left\langle \widehat{N}_F\widehat{N}_{F'}\right\rangle$ is computed using the
2-DM. Electron correlation in this two-electron model can be measured by comparing
the electron fluctuation between $F$ and $F'$ using the 2-DM matrices given in
Fig.~\ref{f:DM2Diag}, \ie
\begin{eqnarray}\label{eq:INDfirst}
I_{\ND}^{\text{MS}} &=&  V_{F,F'}\left[{{^2\DD}^{\HFL}[{^1\DD}]}\right]-
V_{F,F'}\left[{{^2\DD}^{\HFL}[{^1\DD}^{\PHF}]}\right] \\ &=&
%\left\langle \widehat{N}_F\widehat{N}_{F'}\right\rangle^{\Lambda}=
\sum_{ijkl}\sum_{\s\s'} \Lambda^{i^{\s}j^{\s'}}_{k^{\s}l^{\s'}}S^F_{ik}S^{F'}_{jl} \\
\label{eq:IDfirst}
I_{\D}^{\text{MS}} &=& V_{F,F'}\left[{^2\DD}\right] -
V_{F,F'}\left[{{^2\DD}^{\HFL}[{^1\DD}]}\right] \\ &=&
%\left\langle \widehat{N}_F\widehat{N}_{F'}\right\rangle^{\Gamma}=
\sum_{ijkl}\sum_{\s\s'} \Gamma^{i^{\s}j^{\s'}}_{k^{\s}l^{\s'}}S^F_{ik}S^{F'}_{jl}
\end{eqnarray}
where 
$S^{F}_{ik}$ is the overlap between two natural orbitals in the three-dimensional
region of fragment $F$, \ie
\begin{equation} 
S^F_{ik}=\int_F \phi_i^*(\vec{r})\phi_k(\vec{r})d\vec{r}
\end{equation}
Due to symmetry restrictions $S^F_{ii}=S^{F'}_{ii}=1/2$ and it has been proved~\cite{matito:06jce}
that $S^F_{12}S^{F'}_{21}\approx-\frac{1}{4}$. 
Therefore, the formulae can be simplified to
\begin{eqnarray}\label{eq:ID2emb}
I_{\D}^{\text{MS}} &=&
n^{1/2}(1-n)^{1/2}
- 2 n(1-n) \\ \label{eq:IND2emb}
I_{\ND}^{\text{MS}} &=&
2 n(1-n)
\end{eqnarray}
where $n$ is the natural orbital occupancy of either the bonding or antibonding orbital.
These indices are plotted in Fig~\ref{f:2eMB} against the occupation
value. Simple inspection reveals that all the indices are zero for 
full or vanishing occupations, but the dynamic indicator attains larger
values than the nondynamic one in the vicinity of these extreme occupancies.
The nondynamic
indicator is maximal when each electron is equally distributed between the
two orbitals, whereas the dynamic counterpart is minimal in this situation.
The dynamic indicator peaks at $n\approx 0.067$ and $n\approx 0.933$. All
these features are in line with trends that one would expect from dynamic
and nondynamic indicators.\newline

\section{Natural Orbital Formulation}
In order to construct a general natural-orbital based decomposition of electron
correlation in the following we assume that the correlation of a molecule can be
retrieved by summing up the individual contributions of each orbital as calculated
using Eqs.~\ref{eq:ID2emb} and~\ref{eq:IND2emb}, \ie
\begin{eqnarray}\label{eq:IDno}
I_{\D} &=&
\frac{1}{4}\sum_{\s,i}\left[n_i^{\s}(1-n_i^{\s})\right]^{1/2}
-\half\sum_{\s,i} n_i^{\s}(1-n_i^{\s})  \\ \label{eq:INDno}
I_{\ND} &=&
\half\sum_{\s,i} n_i^{\s}(1-n_i^{\s}) 
\end{eqnarray}
where $n_i^{\sigma}$ is the occupation of the spin-natural orbital $i$ with spin $\sigma$.
$I_{\ND}$ coincides with the DFI of the first-order density
matrix, which is a well-known measure
of nondynamic correlation and it is also linked to the number of effectively
unpaired electrons in singlet molecules.~\cite{takatsuka:78tca,staroverov:00cpl,head:03cpl,bochicchio:03cpl,head:03cpl2}
The summation of both quantities gives rise to a total correlation index:
\begin{equation}\label{eq:ITno}
I_{\text{T}}=I_{\D}+I_{\ND}=\frac{1}{4}\sum_{\s,i}\left[n_i^{\s}(1-n_i^{\s})\right]^{1/2}
\end{equation}
Unlike the previous formulation, Eqs.~\ref{eq:decomp}-\ref{eq:gamma}, which depend 
on the 2-DM, the expressions in Eqs.~\ref{eq:IDno}-\ref{eq:ITno} provide simple 
measures of dynamic, nondynamic and total electron correlation in terms of natural orbital
occupancies. Therefore, these expressions are more versatile and can be applied to 
all sort of ab initio methods, density matrix functional theory (DMFT)\cite{piris:14ijqc,pernal:15tcc,mazziotti:12cr}
 and, DFT with
fractional occupancies (or regular DFT by mapping orbital occupancies into Kohn-Sham orbital 
energies\cite{gruning:03jcp}). In addition, these formulae can be naturally decomposed into
orbital contributions and, upon multiplication of orbital amplitudes, they afford
local descriptors of total, dynamic and nondynamic correlation. These descriptors can be
used to measure the local importance of dynamic correlation effects and hence be employed
in the development of local methods.~\cite{zalesny:11book}\newline

$I_{\ND}$ is defined in the interval
$[0,N/2]$, while $I_{\D}$ takes values in $[0,\infty]$ 
because the sum
$\sum_i n_i^{1/2}$ is not bounded above. Indeed, for an extreme case of
infinite occupancies going to zero, the latter sum might diverge. Although
this result is highly unpleasant, it merely has any effect for real systems
where orbital occupancies never reach such situation.\newline

In the L{\"o}wdin-Shull wavefunction the coefficients of the CI expansion take
a very simple form in terms of natural occupancies, $2c_i^2=n_i$. Therefore, one
can easily study these indicators under different electron correlation 
regimes.~\cite{kutzelnigg:99jcp}
A typical dynamic correlation case is characterized by a single dominant configuration,
$c_0\approx1$ (thus $c_i^2\approx0\hspace{0.3cm} \forall i\neq0$). For this system, 
Eq.~\ref{eq:INDno}
goes quickly to zero, and the first term of Eq.~\ref{eq:IDno} dominates over the second
one. Therefore, $I_{\text{T}}\approx I_{\D}$, as one would expect. 
As a prototype of
strong nondynamic correlation effects we analyze the degeneracy between two
electron configurations, which for a two-electron system corresponds to 
$c_0=-c_1=\frac{1}{\sqrt{2}}$ (and $c_i^2=0\hspace{0.3cm} \forall\, i> 1$). In this situation
$I_{\ND}$ attains the maximum value and $I_{\D}$=0, in agreement with
our prediction.

\section{Numerical examples}
We have performed numerical evaluations of the indices for a series
of illustrative systems to validate the formulae derived from the two-electron
model.
FCI calculations with a cc-pVTZ basis set have been performed with a 
modified version of the program of Knowles and Handy~\cite{knowles:89cpc}
for the dissociation of H$_2$ and LiH, for the planar potential energy surface
of H$_4$ and the isoelectronic series of He and Be.
%The cumulant matrix has been obtained with the DMn program.~\cite{dmn} 
Full-valence CASSCF calculations have been performed
with Gaussian09 package~\cite{g09}
in order to obtain Handy's correlation energy decomposition (Eq.~\ref{eq:handy})
for H$_2$ and H$_4$. % and He and Be isoelectronic series.
The same program has been used to obtain CISD/cc-pVTZ
natural orbitals for Be$_2$, CO, F$_2$, HF, Li$_2$, LiH and N$_2$, in order
to calculate $I_{\ND}$ and $I_{\D}$, which are compared against the 
$E_{\ND}$ and $E_{\D}$ values published by Handy.~\cite{mok:96jpc}\newline

\begin{figure}[h!]
\vspace{4.5mm}
\begin{center}
\includegraphics[width=0.37\textwidth]{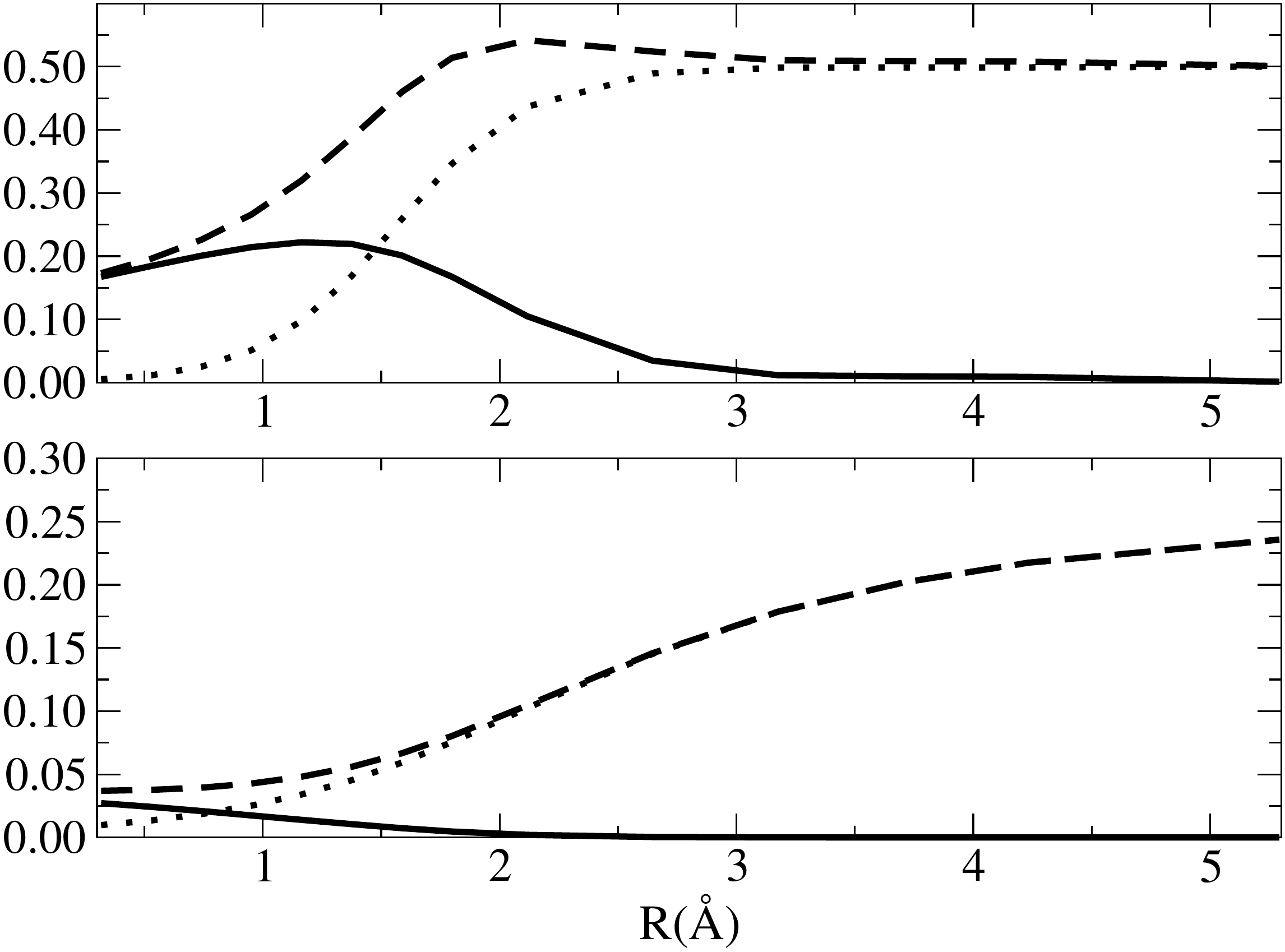}
\end{center}
\caption{$I_{\D}$ (solid line)
$I_{\ND}$ (dotted line) and $I_{\T}$ (dashed line) [\textit{top}], 
$E_{\D}$ (solid line), $E_{\ND}$ and 
$E_{\CORR}$ (dashed line) [\textit{bottom}] 
along the H$_2$ dissociation curve.
The equilibrium distance is 0.7\AA.}
\label{f:H2}
\end{figure}

The values of $I_{\D}$, $I_{\ND}$, $E_{\D}$ and $E_{\ND}$
along the dissociation of H$_2$
are depicted in Fig.~\ref{f:H2}. At the equilibrium distance, the nondynamic correlation
is barely zero and the most important contribution comes from the dynamic correlation.
$I_{\ND}$ presents a sigmoidal growth as the bond stretches and it reaches the maximum 
value of 0.5 at dissociation. $I_{\D}$ slightly increases upon dissociation peaking 
at around 1.2 \AA. From this point on, the index decays to zero because,
as the interaction between electrons decreases, the presence of isolated electrons cannot
give rise to dynamic correlation. 
$E_{\ND}$ and $E_{\D}$ show qualitatively the same trend than $I_{\ND}$ and $E_{\ND}$, 
\ie $E_{\ND}$ dominates over $E_{\D}$ and attains a maximum value at large distances
whereas $E_{\D}$ has larger values than $E_{\ND}$ close to the equilibrium distance and
decays to zero as the H$_2$ molecule stretches.
However, $E_{\D}$ has a maximum at shorter interatomic distance and shows smaller numbers
than $I_{\D}$ (with respect to the total values). We attribute these differences to the
fact that $E_{\D}$ is calculated assuming that a CAS(2,2) calculation of a two-electron
system will include no dynamical correlation. This example puts forward the difficulty
of using reference wavefunctions that do not include dynamic correlation.\newline

\begin{figure}[h!]
\begin{center}
\includegraphics[width=0.44\textwidth]{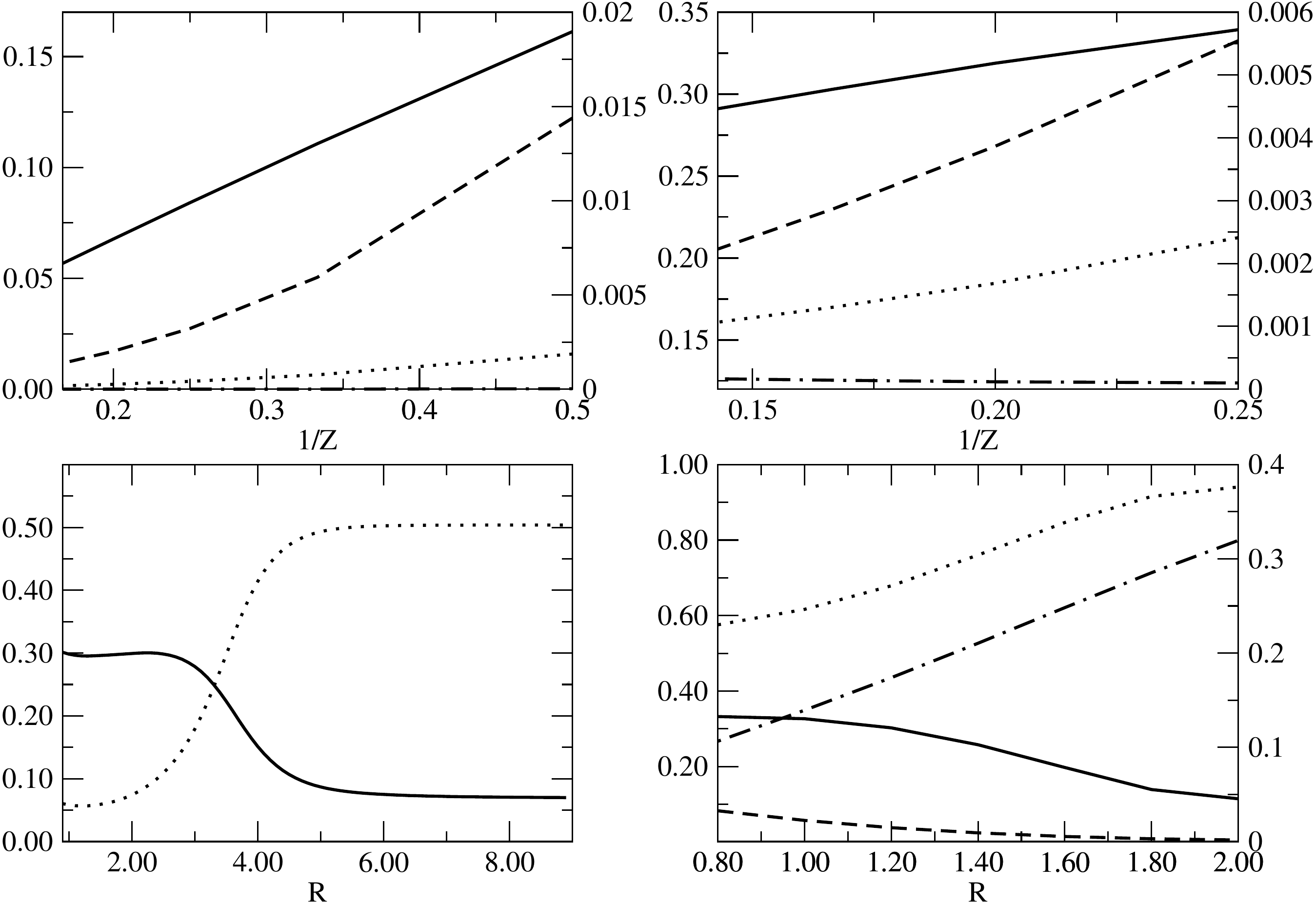}
\end{center}
\caption{From left to right and top to bottom: 
$I_{\D}$ (solid line), $I_{\ND}$ (dotted line), 
$|E_{\D}^{(\text{I})}|/E_{\FCI}$ (dashed line) and $|E_{\ND}^{(\text{I})}|/E_{\FCI}$ 
(dot-dashed line) along 
($i$) the He isoelectronic series as a function of $1/Z$, 
($ii$) the Be isoelectronic series as a function of $1/Z$, ($iii$) the LiH
dissociation as a function of $R_{\text{LiH}}$ (\AA) and 
($iv$) $I_{\D}$ (solid line), $I_{\ND}$ (dotted line), $|E_{\D}^{\HH}|$
(dashed line) and $|E_{\ND}^{\HH}|$ (dot-dashed line) for the dissociation
of the $D_{4h}$ geometry of H$_4$ as we increase the distance between
the H atom and the center of mass, $R$ (\AA).}
\label{f:mols}
\end{figure}

The isolectronic series of Be and He have been widely used to calibrate dynamic and nondynamic
correlation effects.~\cite{valderrama:97jcp,cioslowski:91pra} As the effective nuclear
charge increases the electron correlation decreases because the 
electron distribution concentrates around the nuclei and viceversa. For the He isoelectronic
series the dynamic correlation is far more important than the nondynamic one due to the
absence of orbital degeneracies, as illustrated by Fig.~\ref{f:mols}.
The $2s$ and $2p$ orbitals of Be-like ions are near degenerate
and the HOMO-LUMO gap is thus very small. As $Z$ increases
the gap increases but the relative gap (the gap divided by the average energy of
the orbitals) actually decreases. Gill has called this correlation \textit{type B (nondynamic)
correlation}, whereas the correlation due to the absolute degeneracy of the gap
is referred as \textit{type A (nondynamic) correlation}.~\cite{hollett:11jcp} 
The results in Fig.~\ref{f:mols} show that the DFI can identify 
the nondynamic correlation linked to type A, but it does not
recognize the type B nondynamic correlation. Despite
the presence of degeneracies, $I_{\D}$ is far more important than $I_{\ND}$.
In order to compare these results with a correlation energy decomposition we should
scale $E_{\D}$ and $E_{\ND}$ with respect to $E_{\FCI}$. The energy and,
therefore, the correlation energy are highly
sensible to the external potential. Hence, a crude inspection of $E_{\CORR}$
does not inform about the relative correlation effects in a system with varying
external potential unless an appropriate scaling is performed. Fig.~\ref{f:mols}
contains Cioslowski's decomposition results presented by 
Lude{\~n}a and coworkers,~\cite{valderrama:99jcp} which show similar trends to
$I_{\D}$ and $I_{\ND}$; the most notable exception being the small nondynamic
correlation energy fraction attributed to the Be series.
%~\cite{chakravorty:93pra}\newline
In Fig.~\ref{f:mols} we also include the dissociation of LiH, where
both $I_{\D}$ and $I_{\ND}$ exhibit a similar shape as in H$_2$.\newline

\begin{figure}[h!]
\vspace{5mm}
\begin{center}
\includegraphics[width=0.43\textwidth]{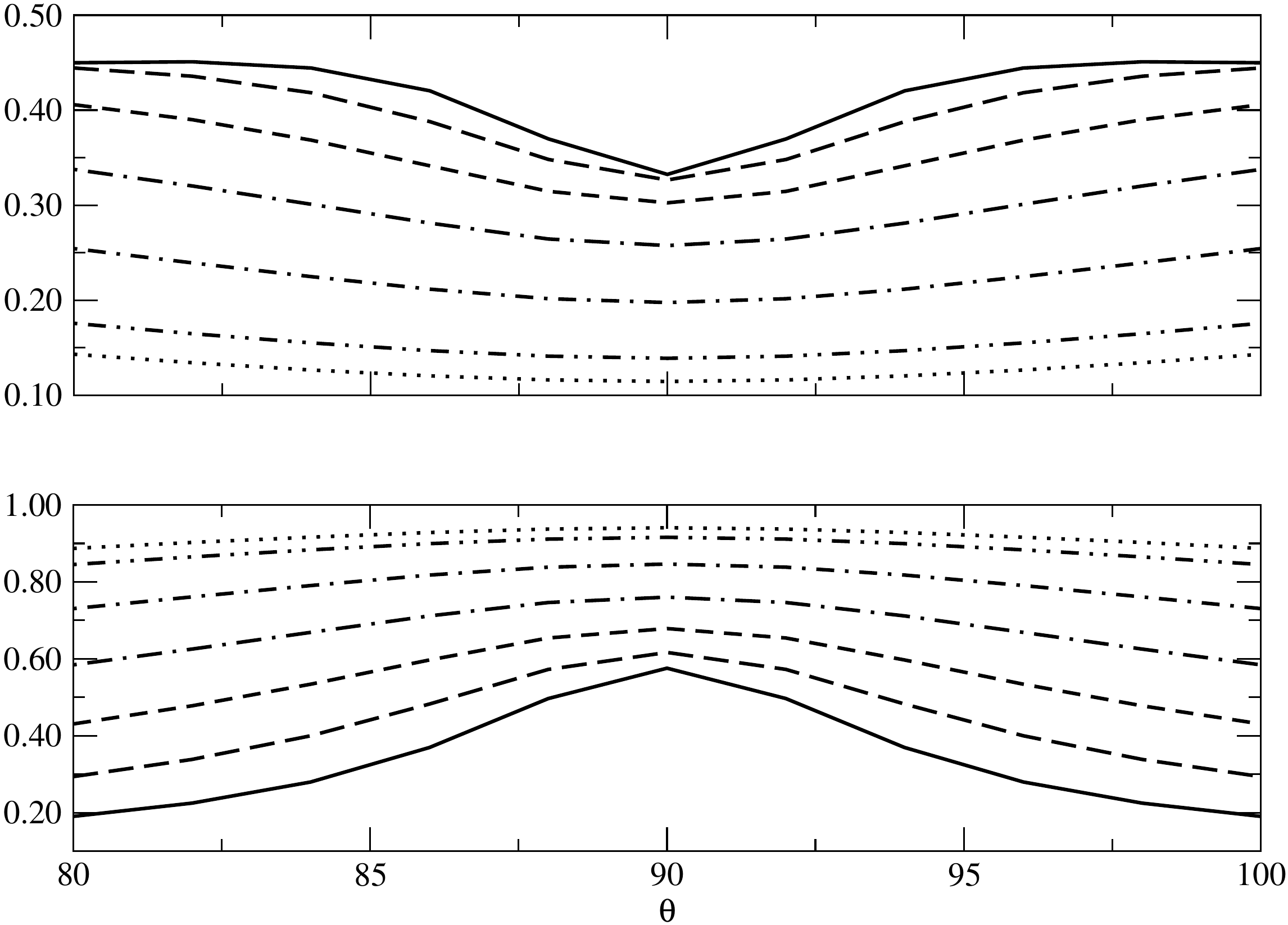}
\end{center}
\caption{$I_{\D}$ (top) and $I_{\ND}$ (bottom) as a function of $\theta$,
the angle between two contiguous H atoms and the center of mass,  for
different $R$. The solid lines represent $R=0.8$\AA$\,$ and the dotted
lines $R=2.0$\AA, the other lines in between correspond to the structures of
$R=1.0, 1.2, 1.4, 1.6$ and $1.8$\AA.}
\label{f:h4}
\end{figure}

\begin{figure}[h!]
\vspace{5mm}
\begin{center}
\includegraphics[width=0.43\textwidth]{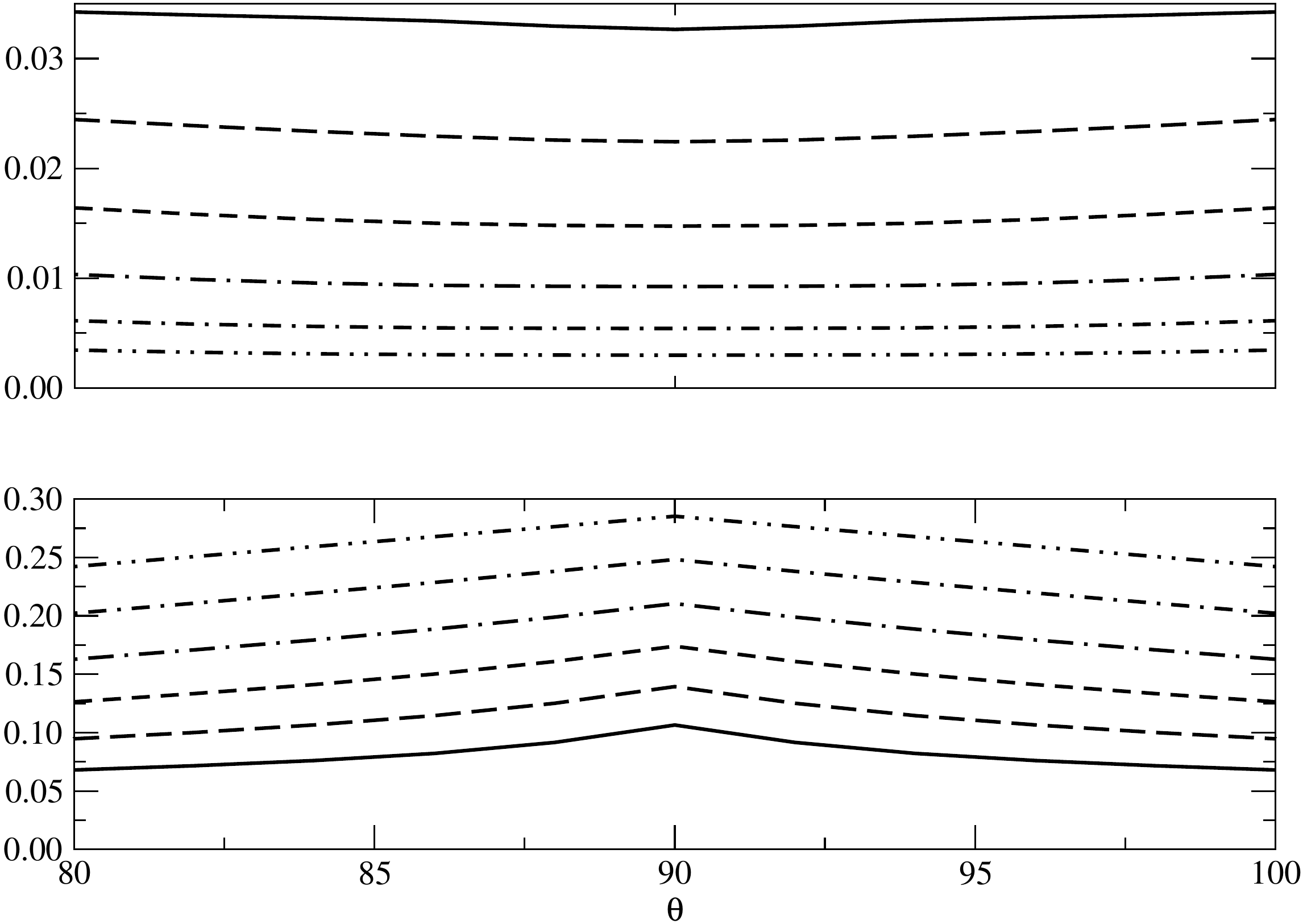}
\end{center}
\caption{$|E_{\D}^{\HH}|$ (top) and $|E_{\ND}^{\HH}|$ (bottom) as a function of $\theta$,
the angle between two contiguous H atoms and the center of mass, for
different $R$. The solid lines represent $R=0.8$\AA$\,$ and the dotted
lines $R=2.0$\AA, the other lines in between correspond to the structures of
$R=1.0, 1.2, 1.4, 1.6$ and $1.8$\AA. Units are hartrees.}
\label{f:h4e}
\end{figure}

In Fig.~\ref{f:mols} we examine the singlet H$_4$ planar potential energy surface~\cite{ramos-cordoba:15jcp} by collecting the $D_{4h}$ structures as a function of $R$,
the distance between H and the center of mass.
These geometries present degenerate $b_{3u}$ and
$b_{2u}$ orbitals that prompt large nondynamic correlation effects as reflected by the
values of $I_{\ND}$ ($|E_{\ND}^{\HH}|$), which are bigger than the $I_{\D}$ ($|E_{\D}^{\HH}|$)
ones for all the 
structures studied. The more stretched the H$_4$ structure, the larger the nondynamic 
correlation effects, whereas dynamic correlation is most notorious for constricted
geometries. The transition from $D_{4h}$ to $D_{2h}$ structures (that is controled
by the angle $\theta$ between two contiguous H atoms and the center of mass)
can be used to see the change from nondynamic-controled region (near the $D_{4h}$ 
geometries) to a region of predominant dynamic correlation. 
As $R$ increases, the values of $I_{\D}$ and $|E_{\D}^{\HH}|$ decrease and
the plots as a function of $\theta$ show a minimum at the $D_{4h}$ structure and 
become flatter, in agreement with the small dynamic correlation
expected at stretched geometries (see Figs.~\ref{f:h4} and~\ref{f:h4e}). $I_{\ND}$ 
and $|E_{\ND}^{\HH}|$ show exactly the opposite
profiles, with largest values at the D$_{4h}$ geometries and flatter profiles for large $R$.\newline

\begin{figure}[h!]
 \centering
 \includegraphics[scale=0.5]{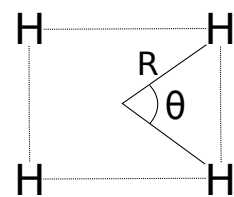}
 \caption{The D$_{4h}$/D$_{2h}$ potential energy surface of \hh is given in terms of two
coordinates: $R$ and $\theta$.}
 \label{f:h4mol}
\end{figure}


Finally, Table~\ref{tbl:newmols} collects a few diatomic molecules at the
equilibrium distance ($R_e$) and at a stretched geometry ($1.5R_e$).
As expected, the nondynamic correlation indices increase
upon bond length elongation
with only a few exceptions. Be$_2$ is an exception
due to the unusually large nondynamic correlation nature at the ground state
distance, as
confirmed by the large $I_{\ND}$ value at equilibrium that barely changes
for the stretched bond length. LiH and Li$_2$ show a $I_{\ND}$ reduction
upon stretching the internuclear distance, whereas $|E_{\ND}^{\HH}|$ 
remains constant. This feature was recognized by Handy and coworkers, 
who suggested an alternative electron correlation energy partition,
consisting of a CASSCF calculation with an active space such that the angular
correlation is not incorporated.~\cite{mok:96jpc}
The results obtained with this alternative
partition do
show some increase of $|E_{\ND}^{\HH}|$ for LiH and Li$_2$ 
due to bond length elongation,~\cite{mok:96jpc}
however, they put forward the limitation of the CASSCF wavefunction as a 
reference containing no dynamic correlation.

\begin{table}[]
\centering
\small{
\caption{CISD/cc-pVTZ $I_{\D}$, $I_{\ND}$, $E_{\D}$ and $E_{\ND}$ values for a series
of diatomic molecules at its ground state geometry ($R_e$) and at the
$1.5R_e$ interatomic separation. $I_{\D}$, $I_{\ND}$ units are electrons
and $E_{\D}$ and $E_{\ND}$ units are hartrees.
$^{a}\,E_{\D}$ and $E_{\ND}$ are taken
Ref.~\citenum{mok:96jpc}. }
\label{tbl:newmols}
\begin{tabular}{ccccccccc}\hline
&  \multicolumn{2}{c}{$I_{\D}$} & \multicolumn{2}{c}{$I_{\ND}$} 
&  \multicolumn{2}{c}{$E_{\D}$$^{a}$} & \multicolumn{2}{c}{$E_{\ND}$$^{a}$}      \\ \hline
& $R_e$ & $1.5R_e$ & $R_e$ & $1.5R_e$ & $R_e$ & $1.5R_e$ & $R_e$ & $1.5R_e$ \\
\hline
Be$_2$ & 0.60  &  0.57 & 0.29  & 0.27 & 0.12  &  0.11 & 0.09  &  0.09 \\
CO  & 0.83  &  0.89 & 0.17  &  0.24 & 0.41  &  0.41 & 0.13  &  0.22 \\
F$_2$  & 0.91  &  0.89 & 0.16  &  0.29 & 0.69  &  0.66 & 0.08  &  0.22 \\
HF  & 0.55  &  0.57 & 0.09  &  0.12 & 0.37  &  0.37 & 0.02  &  0.05 \\
Li$_2$ & 0.31  &  0.28 & 0.18  &  0.24 & 0.10  &  0.10 & 0.03  &  0.03 \\
LiH & 0.23  &  0.24 & 0.06  &  0.10 & 0.06  &  0.06 & 0.03  &  0.04 \\
N$_2$  & 0.84  &  0.91 & 0.18  &  0.29 & 0.41  &  0.41 & 0.15  &  0.31 \\ \hline 
\end{tabular}
}
\end{table}

\section{Conclusions}
In this paper we have developed physically-sound electron correlation matrices
that account for total (Eq.~\ref{eq:decomp}), nondynamic (Eq.~\ref{eq:lambda}) and
dynamic (Eq.~\ref{eq:gamma}) correlation effects and depend on the pair density. These
expressions are applied to a minimal-basis two-fragment two-electron model at different 
inter-fragment separations to afford dynamic (Eq.~\ref{eq:IDno}), nondynamic (Eq.~\ref{eq:INDno}) and
total (Eq.~\ref{eq:ITno}) electron correlation indicators in terms of natural orbital
occupancies. 
Unlike other existing indicators of electron correlation, they neither depend upon the definition
of wavefunctions containing only dynamic or nondynamic correlation effects, nor they need
the calculation of the exact wavefunctions. The indicators here developed do not rely
on reference wavefunctions and can be applied to any method to analyze the effect of
dynamic and nondynamic correlation provided that natural orbital occupancies
are available.
The electron correlation indicators have been analzed in a set of representative examples
providing different dynamic and nondynamic electron correlation regimes. Results are also
compared against existing electron correlation measures, demonstrating the validity
of the new expressions.\newline

The latter expressions can be applied to 
all sort of ab initio methods, DMFT and, ensemble DFT with fractional occupancies~\cite{fromager:15mp}
(or DFT by mapping orbital occupancies into Kohn-Sham orbital energies\cite{gruning:03jcp}). 
In addition, these formulae can be naturally decomposed into
orbital contributions and, upon multiplication of orbital amplitudes, they afford
local descriptors of total, dynamic and nondynamic correlation. These descriptors 
measure the local importance of dynamic correlation and hence can be used
in the development of local methods.\cite{zalesny:11book} 

\section{Acknowledgements}
This research has been funded by the Spanish MINECO
Projects No. CTQ2014-52525-P and CTQ2014-59212-P, and the Basque 
Country Consolidated Group Project No. IT588-13.
ERC acknowledges funding from the European Union's 
Horizon 2020 research and innovation programme under the Marie 
Sklodowska-Curie grant agreement No 660943.
E.M. acknowledges fruitful discussions with Drs. Savin, Szabados and Cioslowski.

\providecommand*{\mcitethebibliography}{\thebibliography}
\csname @ifundefined\endcsname{endmcitethebibliography}
{\let\endmcitethebibliography\endthebibliography}{}


\begin{mcitethebibliography}{76}
\providecommand*{\natexlab}[1]{#1}
\providecommand*{\mciteSetBstSublistMode}[1]{}
\providecommand*{\mciteSetBstMaxWidthForm}[2]{}
\providecommand*{\mciteBstWouldAddEndPuncttrue}
  {\def\EndOfBibitem{\unskip.}}
\providecommand*{\mciteBstWouldAddEndPunctfalse}
  {\let\EndOfBibitem\relax}
\providecommand*{\mciteSetBstMidEndSepPunct}[3]{}
\providecommand*{\mciteSetBstSublistLabelBeginEnd}[3]{}
\providecommand*{\EndOfBibitem}{}
\mciteSetBstSublistMode{f}
\mciteSetBstMaxWidthForm{subitem}
{(\emph{\alph{mcitesubitemcount}})}
\mciteSetBstSublistLabelBeginEnd{\mcitemaxwidthsubitemform\space}
{\relax}{\relax}

\bibitem[H{\"a}ttig \emph{et~al.}(2012)H{\"a}ttig, Klopper, K{\"o}hn, and
  Tew]{hattig:12cr}
C.~H{\"a}ttig, W.~Klopper, A.~K{\"o}hn and D.~P. Tew, \emph{Chem. Rev.}, 2012,
  \textbf{112}, 4--74\relax
\mciteBstWouldAddEndPuncttrue
\mciteSetBstMidEndSepPunct{\mcitedefaultmidpunct}
{\mcitedefaultendpunct}{\mcitedefaultseppunct}\relax
\EndOfBibitem
\bibitem[Cremer(2001)]{cremer:01mp}
D.~Cremer, \emph{Molec. Phys.}, 2001, \textbf{99}, 1899--1940\relax
\mciteBstWouldAddEndPuncttrue
\mciteSetBstMidEndSepPunct{\mcitedefaultmidpunct}
{\mcitedefaultendpunct}{\mcitedefaultseppunct}\relax
\EndOfBibitem
\bibitem[Ziesche(2000)]{ziesche:00the}
P.~Ziesche, \emph{J. Mol. Struct. (Theochem)}, 2000, \textbf{527}, 35--50\relax
\mciteBstWouldAddEndPuncttrue
\mciteSetBstMidEndSepPunct{\mcitedefaultmidpunct}
{\mcitedefaultendpunct}{\mcitedefaultseppunct}\relax
\EndOfBibitem
\bibitem[Tew \emph{et~al.}(2007)Tew, Klopper, and Helgaker]{tew:07jcc}
D.~P. Tew, W.~Klopper and T.~Helgaker, \emph{J. Comput. Chem.}, 2007,
  \textbf{28}, 1307--1320\relax
\mciteBstWouldAddEndPuncttrue
\mciteSetBstMidEndSepPunct{\mcitedefaultmidpunct}
{\mcitedefaultendpunct}{\mcitedefaultseppunct}\relax
\EndOfBibitem
\bibitem[L{\"o}wdin(1955)]{lowdin:55pr}
P.-O. L{\"o}wdin, \emph{Phys. Rev.}, 1955, \textbf{97}, 1474--1489\relax
\mciteBstWouldAddEndPuncttrue
\mciteSetBstMidEndSepPunct{\mcitedefaultmidpunct}
{\mcitedefaultendpunct}{\mcitedefaultseppunct}\relax
\EndOfBibitem
\bibitem[Kutzelnigg \emph{et~al.}(1968)Kutzelnigg, Del~Re, and
  Berthier]{kutzelnigg:68pr}
W.~Kutzelnigg, G.~Del~Re and G.~Berthier, \emph{Phys. Rev.}, 1968,
  \textbf{172}, 49\relax
\mciteBstWouldAddEndPuncttrue
\mciteSetBstMidEndSepPunct{\mcitedefaultmidpunct}
{\mcitedefaultendpunct}{\mcitedefaultseppunct}\relax
\EndOfBibitem
\bibitem[Raghavachari and Anderson(1996)]{raghavachari:96jpc}
K.~Raghavachari and J.~B. Anderson, \emph{J. Phys. Chem.}, 1996, \textbf{100},
  12960--12973\relax
\mciteBstWouldAddEndPuncttrue
\mciteSetBstMidEndSepPunct{\mcitedefaultmidpunct}
{\mcitedefaultendpunct}{\mcitedefaultseppunct}\relax
\EndOfBibitem
\bibitem[Juh{\'a}sz and Mazziotti(2006)]{juhasz:06jcp}
T.~Juh{\'a}sz and D.~A. Mazziotti, \emph{J. Chem. Phys.}, 2006, \textbf{125},
  174105\relax
\mciteBstWouldAddEndPuncttrue
\mciteSetBstMidEndSepPunct{\mcitedefaultmidpunct}
{\mcitedefaultendpunct}{\mcitedefaultseppunct}\relax
\EndOfBibitem
\bibitem[Alcoba \emph{et~al.}(2010)Alcoba, Bochicchio, Lain, and
  Torre]{alcoba:10jcp}
D.~Alcoba, R.~Bochicchio, L.~Lain and A.~Torre, \emph{J. Chem. Phys.}, 2010,
  \textbf{133}, 144104\relax
\mciteBstWouldAddEndPuncttrue
\mciteSetBstMidEndSepPunct{\mcitedefaultmidpunct}
{\mcitedefaultendpunct}{\mcitedefaultseppunct}\relax
\EndOfBibitem
\bibitem[Skolnik and Mazziotti(2013)]{skolnik:13pra}
J.~T. Skolnik and D.~A. Mazziotti, \emph{Phys. Rev. A}, 2013, \textbf{88},
  032517\relax
\mciteBstWouldAddEndPuncttrue
\mciteSetBstMidEndSepPunct{\mcitedefaultmidpunct}
{\mcitedefaultendpunct}{\mcitedefaultseppunct}\relax
\EndOfBibitem
\bibitem[Sinano{\u{g}}lu(1964)]{sinanoglu:64acp}
O.~Sinano{\u{g}}lu, \emph{Adv. Chem. Phys.}, 1964, \textbf{6}, 315--412\relax
\mciteBstWouldAddEndPuncttrue
\mciteSetBstMidEndSepPunct{\mcitedefaultmidpunct}
{\mcitedefaultendpunct}{\mcitedefaultseppunct}\relax
\EndOfBibitem
\bibitem[Bartlett and Stanton(2007)]{bartlett:07rcc}
R.~J. Bartlett and J.~F. Stanton, \emph{Rev. Comp. Chem.}, 2007, \textbf{5},
  65--169\relax
\mciteBstWouldAddEndPuncttrue
\mciteSetBstMidEndSepPunct{\mcitedefaultmidpunct}
{\mcitedefaultendpunct}{\mcitedefaultseppunct}\relax
\EndOfBibitem
\bibitem[Lennard-Jones(1952)]{lennardjones:52jcp}
J.~E. Lennard-Jones, \emph{J. Chem. Phys.}, 1952, \textbf{20}, 1024\relax
\mciteBstWouldAddEndPuncttrue
\mciteSetBstMidEndSepPunct{\mcitedefaultmidpunct}
{\mcitedefaultendpunct}{\mcitedefaultseppunct}\relax
\EndOfBibitem
\bibitem[Lie and Clementi(1974)]{lie:74jcp}
G.~C. Lie and E.~Clementi, \emph{J. Chem. Phys.}, 1974, \textbf{60},
  1275--1287\relax
\mciteBstWouldAddEndPuncttrue
\mciteSetBstMidEndSepPunct{\mcitedefaultmidpunct}
{\mcitedefaultendpunct}{\mcitedefaultseppunct}\relax
\EndOfBibitem
\bibitem[Lee \emph{et~al.}(1989)Lee, Rice, Scuseria, and Schaefer]{lee:89tca}
T.~J. Lee, J.~E. Rice, G.~E. Scuseria and H.~F. Schaefer, \emph{Theor. Chim.
  Acta (Berlin)}, 1989, \textbf{75}, 81\relax
\mciteBstWouldAddEndPuncttrue
\mciteSetBstMidEndSepPunct{\mcitedefaultmidpunct}
{\mcitedefaultendpunct}{\mcitedefaultseppunct}\relax
\EndOfBibitem
\bibitem[Lee and Taylor(1989)]{lee:89ijqc}
T.~J. Lee and P.~R. Taylor, \emph{Int. J. Quant. Chem.}, 1989, \textbf{23},
  199--207\relax
\mciteBstWouldAddEndPuncttrue
\mciteSetBstMidEndSepPunct{\mcitedefaultmidpunct}
{\mcitedefaultendpunct}{\mcitedefaultseppunct}\relax
\EndOfBibitem
\bibitem[Janssen and Nielsen(1998)]{janssen:98cpl}
C.~L. Janssen and I.~M.~B. Nielsen, \emph{Chem. Phys. Lett.}, 1998,
  \textbf{290}, 290\relax
\mciteBstWouldAddEndPuncttrue
\mciteSetBstMidEndSepPunct{\mcitedefaultmidpunct}
{\mcitedefaultendpunct}{\mcitedefaultseppunct}\relax
\EndOfBibitem
\bibitem[Lee(2003)]{lee:03cpl}
T.~J. Lee, \emph{Chem. Phys. Lett.}, 2003, \textbf{372}, 362--367\relax
\mciteBstWouldAddEndPuncttrue
\mciteSetBstMidEndSepPunct{\mcitedefaultmidpunct}
{\mcitedefaultendpunct}{\mcitedefaultseppunct}\relax
\EndOfBibitem
\bibitem[Cioslowski(1992)]{cioslowski:92tca}
J.~Cioslowski, \emph{Theor. Chim. Acta (Berlin)}, 1992, \textbf{81},
  319--327\relax
\mciteBstWouldAddEndPuncttrue
\mciteSetBstMidEndSepPunct{\mcitedefaultmidpunct}
{\mcitedefaultendpunct}{\mcitedefaultseppunct}\relax
\EndOfBibitem
\bibitem[Andersson \emph{et~al.}(1992)Andersson, Malmqvist, and
  Roos]{andersson:92jcp}
K.~Andersson, P.-{\AA}. Malmqvist and B.~O. Roos, \emph{J. Chem. Phys.}, 1992,
  \textbf{96}, 1218--1226\relax
\mciteBstWouldAddEndPuncttrue
\mciteSetBstMidEndSepPunct{\mcitedefaultmidpunct}
{\mcitedefaultendpunct}{\mcitedefaultseppunct}\relax
\EndOfBibitem
\bibitem[Buenker \emph{et~al.}(1978)Buenker, Peyerimhoff, and
  Butscher]{buenker:78mp}
R.~J. Buenker, S.~D. Peyerimhoff and W.~Butscher, \emph{Molec. Phys.}, 1978,
  \textbf{35}, 771--791\relax
\mciteBstWouldAddEndPuncttrue
\mciteSetBstMidEndSepPunct{\mcitedefaultmidpunct}
{\mcitedefaultendpunct}{\mcitedefaultseppunct}\relax
\EndOfBibitem
\bibitem[Mazziotti(2007)]{mazziotti:07pra}
D.~A. Mazziotti, \emph{Phys. Rev. A}, 2007, \textbf{76}, 052502\relax
\mciteBstWouldAddEndPuncttrue
\mciteSetBstMidEndSepPunct{\mcitedefaultmidpunct}
{\mcitedefaultendpunct}{\mcitedefaultseppunct}\relax
\EndOfBibitem
\bibitem[Snyder~Jr and Mazziotti(2012)]{snyder:12pccp}
J.~W. Snyder~Jr and D.~A. Mazziotti, \emph{Phys. Chem. Chem. Phys.}, 2012,
  \textbf{14}, 1660--1667\relax
\mciteBstWouldAddEndPuncttrue
\mciteSetBstMidEndSepPunct{\mcitedefaultmidpunct}
{\mcitedefaultendpunct}{\mcitedefaultseppunct}\relax
\EndOfBibitem
\bibitem[Savin(1988)]{savin:88ijqc}
A.~Savin, \emph{Int. J. Quant. Chem.}, 1988, \textbf{34}, 59--69\relax
\mciteBstWouldAddEndPuncttrue
\mciteSetBstMidEndSepPunct{\mcitedefaultmidpunct}
{\mcitedefaultendpunct}{\mcitedefaultseppunct}\relax
\EndOfBibitem
\bibitem[Savin(1991)]{savin:91chapter}
A.~Savin, \emph{Density functional methods in chemistry}, Springer, 1991, pp.
  213--230\relax
\mciteBstWouldAddEndPuncttrue
\mciteSetBstMidEndSepPunct{\mcitedefaultmidpunct}
{\mcitedefaultendpunct}{\mcitedefaultseppunct}\relax
\EndOfBibitem
\bibitem[Savin(1996)]{savin:96bc}
A.~Savin, \emph{On degeneracy, near-degeneracy and density functional theory.
  In Recent Developments of Modern Density Functional Theory}, Elsevier:
  Amsterdam, 1996\relax
\mciteBstWouldAddEndPuncttrue
\mciteSetBstMidEndSepPunct{\mcitedefaultmidpunct}
{\mcitedefaultendpunct}{\mcitedefaultseppunct}\relax
\EndOfBibitem
\bibitem[Jaramillo \emph{et~al.}(2003)Jaramillo, Scuseria, and
  Ernzerhof]{jaramillo:03jcp}
J.~Jaramillo, G.~E. Scuseria and M.~Ernzerhof, \emph{J. Chem. Phys.}, 2003,
  \textbf{118}, 1068--1073\relax
\mciteBstWouldAddEndPuncttrue
\mciteSetBstMidEndSepPunct{\mcitedefaultmidpunct}
{\mcitedefaultendpunct}{\mcitedefaultseppunct}\relax
\EndOfBibitem
\bibitem[Becke(1993)]{becke:93jcp}
A.~D. Becke, \emph{J. Chem. Phys.}, 1993, \textbf{98}, 5648--5652\relax
\mciteBstWouldAddEndPuncttrue
\mciteSetBstMidEndSepPunct{\mcitedefaultmidpunct}
{\mcitedefaultendpunct}{\mcitedefaultseppunct}\relax
\EndOfBibitem
\bibitem[Becke(1993)]{becke:93jcp1}
A.~D. Becke, \emph{J. Chem. Phys.}, 1993, \textbf{98}, 1372--1377\relax
\mciteBstWouldAddEndPuncttrue
\mciteSetBstMidEndSepPunct{\mcitedefaultmidpunct}
{\mcitedefaultendpunct}{\mcitedefaultseppunct}\relax
\EndOfBibitem
\bibitem[Perdew \emph{et~al.}(1996)Perdew, Ernzerhof, and Burke]{perdew:96jcp}
J.~P. Perdew, M.~Ernzerhof and K.~Burke, \emph{J. Chem. Phys.}, 1996,
  \textbf{105}, 9982--9985\relax
\mciteBstWouldAddEndPuncttrue
\mciteSetBstMidEndSepPunct{\mcitedefaultmidpunct}
{\mcitedefaultendpunct}{\mcitedefaultseppunct}\relax
\EndOfBibitem
\bibitem[Cruz \emph{et~al.}(1998)Cruz, Lam, and Burke]{cruz:98jpca}
F.~G. Cruz, K.-C. Lam and K.~Burke, \emph{J. Phys. Chem. A}, 1998,
  \textbf{102}, 4911--4917\relax
\mciteBstWouldAddEndPuncttrue
\mciteSetBstMidEndSepPunct{\mcitedefaultmidpunct}
{\mcitedefaultendpunct}{\mcitedefaultseppunct}\relax
\EndOfBibitem
\bibitem[Iikura \emph{et~al.}(2001)Iikura, Tsuneda, Yanai, and
  Hirao]{iikura:01jcp}
H.~Iikura, T.~Tsuneda, T.~Yanai and K.~Hirao, \emph{J. Chem. Phys.}, 2001,
  \textbf{115}, 3540--3544\relax
\mciteBstWouldAddEndPuncttrue
\mciteSetBstMidEndSepPunct{\mcitedefaultmidpunct}
{\mcitedefaultendpunct}{\mcitedefaultseppunct}\relax
\EndOfBibitem
\bibitem[Baer \emph{et~al.}(2009)Baer, Livshits, and Salzner]{baer:09arpc}
R.~Baer, E.~Livshits and U.~Salzner, \emph{Ann. Rev. Phys. Chem.}, 2009,
  \textbf{61}, 85\relax
\mciteBstWouldAddEndPuncttrue
\mciteSetBstMidEndSepPunct{\mcitedefaultmidpunct}
{\mcitedefaultendpunct}{\mcitedefaultseppunct}\relax
\EndOfBibitem
\bibitem[Arbuznikov and Kaupp(2014)]{arbuznikov:14jcp}
A.~V. Arbuznikov and M.~Kaupp, \emph{J. Chem. Phys.}, 2014, \textbf{141},
  204101\relax
\mciteBstWouldAddEndPuncttrue
\mciteSetBstMidEndSepPunct{\mcitedefaultmidpunct}
{\mcitedefaultendpunct}{\mcitedefaultseppunct}\relax
\EndOfBibitem
\bibitem[Henderson \emph{et~al.}(2008)Henderson, Janesko, and
  Scuseria]{henderson:08jpca}
T.~M. Henderson, B.~G. Janesko and G.~E. Scuseria, \emph{J. Phys. Chem. A},
  2008, \textbf{112}, 12530--12542\relax
\mciteBstWouldAddEndPuncttrue
\mciteSetBstMidEndSepPunct{\mcitedefaultmidpunct}
{\mcitedefaultendpunct}{\mcitedefaultseppunct}\relax
\EndOfBibitem
\bibitem[Johnson(2014)]{johnson:14jcp}
E.~R. Johnson, \emph{J. Chem. Phys.}, 2014, \textbf{141}, 124120\relax
\mciteBstWouldAddEndPuncttrue
\mciteSetBstMidEndSepPunct{\mcitedefaultmidpunct}
{\mcitedefaultendpunct}{\mcitedefaultseppunct}\relax
\EndOfBibitem
\bibitem[Becke(1988)]{becke:88jcp}
A.~D. Becke, \emph{J. Chem. Phys.}, 1988, \textbf{88}, 2547--2553\relax
\mciteBstWouldAddEndPuncttrue
\mciteSetBstMidEndSepPunct{\mcitedefaultmidpunct}
{\mcitedefaultendpunct}{\mcitedefaultseppunct}\relax
\EndOfBibitem
\bibitem[De~Silva and Corminboeuf(2015)]{silva:15jcp}
P.~De~Silva and C.~Corminboeuf, \emph{J. Chem. Phys.}, 2015, \textbf{142},
  074112\relax
\mciteBstWouldAddEndPuncttrue
\mciteSetBstMidEndSepPunct{\mcitedefaultmidpunct}
{\mcitedefaultendpunct}{\mcitedefaultseppunct}\relax
\EndOfBibitem
\bibitem[Grimme and Hansen(2015)]{grimme:15ang}
S.~Grimme and A.~Hansen, \emph{Angew. Chem. Int. Ed. Engl.}, 2015, \textbf{54},
  12308--12313\relax
\mciteBstWouldAddEndPuncttrue
\mciteSetBstMidEndSepPunct{\mcitedefaultmidpunct}
{\mcitedefaultendpunct}{\mcitedefaultseppunct}\relax
\EndOfBibitem
\bibitem[Boguslawski \emph{et~al.}(2012)Boguslawski, Tecmer, Legeza, and
  Reiher]{boguslawski:12jpcl}
K.~Boguslawski, P.~Tecmer, O.~Legeza and M.~Reiher, \emph{J. Phys. Chem.
  Lett.}, 2012, \textbf{3}, 3129--3135\relax
\mciteBstWouldAddEndPuncttrue
\mciteSetBstMidEndSepPunct{\mcitedefaultmidpunct}
{\mcitedefaultendpunct}{\mcitedefaultseppunct}\relax
\EndOfBibitem
\bibitem[Raeber and Mazziotti(2015)]{raeber:15pra}
A.~Raeber and D.~A. Mazziotti, \emph{Phys. Rev. A}, 2015, \textbf{92},
  052502\relax
\mciteBstWouldAddEndPuncttrue
\mciteSetBstMidEndSepPunct{\mcitedefaultmidpunct}
{\mcitedefaultendpunct}{\mcitedefaultseppunct}\relax
\EndOfBibitem
\bibitem[Pulay(1983)]{pulay:83cpl}
P.~Pulay, \emph{Chem. Phys. Lett.}, 1983, \textbf{100}, 151--154\relax
\mciteBstWouldAddEndPuncttrue
\mciteSetBstMidEndSepPunct{\mcitedefaultmidpunct}
{\mcitedefaultendpunct}{\mcitedefaultseppunct}\relax
\EndOfBibitem
\bibitem[Zalesny \emph{et~al.}(2011)Zalesny, Papadopoulos, Mezey, and
  Leszczynski]{zalesny:11book}
R.~Zalesny, M.~G. Papadopoulos, P.~G. Mezey and J.~Leszczynski,
  \emph{Linear-scaling techniques in computational chemistry and physics:
  Methods and applications}, Springer Science+ Business Media BV, 2011\relax
\mciteBstWouldAddEndPuncttrue
\mciteSetBstMidEndSepPunct{\mcitedefaultmidpunct}
{\mcitedefaultendpunct}{\mcitedefaultseppunct}\relax
\EndOfBibitem
\bibitem[sha(1984)]{shavitt:84chapter}
\emph{Advanced Theories and Computational Approaches to the Electronic
  Structure of Molecules}, ed. C.~E. D.~e. Thom. H. Dunning~Jr., Raymond A.
  Bair~(auth.), Springer Netherlands, 1984, pp. 185--196\relax
\mciteBstWouldAddEndPuncttrue
\mciteSetBstMidEndSepPunct{\mcitedefaultmidpunct}
{\mcitedefaultendpunct}{\mcitedefaultseppunct}\relax
\EndOfBibitem
\bibitem[Cioslowski(1991)]{cioslowski:91pra}
J.~Cioslowski, \emph{Phys. Rev. A}, 1991, \textbf{43}, 1223--1228\relax
\mciteBstWouldAddEndPuncttrue
\mciteSetBstMidEndSepPunct{\mcitedefaultmidpunct}
{\mcitedefaultendpunct}{\mcitedefaultseppunct}\relax
\EndOfBibitem
\bibitem[Valderrama \emph{et~al.}(1997)Valderrama, Lude{\~n}a, and
  Hinze]{valderrama:97jcp}
E.~Valderrama, E.~V. Lude{\~n}a and J.~Hinze, \emph{J. Chem. Phys.}, 1997,
  \textbf{106}, 9227--9235\relax
\mciteBstWouldAddEndPuncttrue
\mciteSetBstMidEndSepPunct{\mcitedefaultmidpunct}
{\mcitedefaultendpunct}{\mcitedefaultseppunct}\relax
\EndOfBibitem
\bibitem[Valderrama \emph{et~al.}(1999)Valderrama, Lude{\~n}a, and
  Hinze]{valderrama:99jcp}
E.~Valderrama, E.~V. Lude{\~n}a and J.~Hinze, \emph{J. Chem. Phys.}, 1999,
  \textbf{110}, 2343--2353\relax
\mciteBstWouldAddEndPuncttrue
\mciteSetBstMidEndSepPunct{\mcitedefaultmidpunct}
{\mcitedefaultendpunct}{\mcitedefaultseppunct}\relax
\EndOfBibitem
\bibitem[Mok \emph{et~al.}(1996)Mok, Neumann, and Handy]{mok:96jpc}
D.~K.~W. Mok, R.~Neumann and N.~C. Handy, \emph{J. Phys. Chem.}, 1996,
  \textbf{100}, 6225--6230\relax
\mciteBstWouldAddEndPuncttrue
\mciteSetBstMidEndSepPunct{\mcitedefaultmidpunct}
{\mcitedefaultendpunct}{\mcitedefaultseppunct}\relax
\EndOfBibitem
\bibitem[Davidson \emph{et~al.}(1991)Davidson, Hagstrom, Chakravorty, Umar, and
  Fischer]{davidson:91pra}
E.~Davidson, S.~Hagstrom, S.~Chakravorty, V.~Umar and C.~Fischer, \emph{Phys.
  Rev. A}, 1991, \textbf{44}, 7071\relax
\mciteBstWouldAddEndPuncttrue
\mciteSetBstMidEndSepPunct{\mcitedefaultmidpunct}
{\mcitedefaultendpunct}{\mcitedefaultseppunct}\relax
\EndOfBibitem
\bibitem[Valderrama \emph{et~al.}(2001)Valderrama, Mercero, and
  Ugalde]{valderrama:01jpb}
E.~Valderrama, J.~Mercero and J.~Ugalde, \emph{J. Phys. B: Atom. Molec. Phys.},
  2001, \textbf{34}, 275\relax
\mciteBstWouldAddEndPuncttrue
\mciteSetBstMidEndSepPunct{\mcitedefaultmidpunct}
{\mcitedefaultendpunct}{\mcitedefaultseppunct}\relax
\EndOfBibitem
\bibitem[Piris and Ugalde(2014)]{piris:14ijqc}
M.~Piris and J.~Ugalde, \emph{Int. J. Quant. Chem.}, 2014, \textbf{114},
  1169--1175\relax
\mciteBstWouldAddEndPuncttrue
\mciteSetBstMidEndSepPunct{\mcitedefaultmidpunct}
{\mcitedefaultendpunct}{\mcitedefaultseppunct}\relax
\EndOfBibitem
\bibitem[Pernal and Giesbertz(2015)]{pernal:15tcc}
K.~Pernal and K.~J.~H. Giesbertz, \emph{Top. Curr. Chem.}, 2015, \textbf{368},
  125\relax
\mciteBstWouldAddEndPuncttrue
\mciteSetBstMidEndSepPunct{\mcitedefaultmidpunct}
{\mcitedefaultendpunct}{\mcitedefaultseppunct}\relax
\EndOfBibitem
\bibitem[Gr{\"u}ning \emph{et~al.}(2003)Gr{\"u}ning, Gritsenko, and
  Baerends]{gruning:03jcp}
M.~Gr{\"u}ning, O.~Gritsenko and E.~Baerends, \emph{J. Chem. Phys.}, 2003,
  \textbf{118}, 7183--7192\relax
\mciteBstWouldAddEndPuncttrue
\mciteSetBstMidEndSepPunct{\mcitedefaultmidpunct}
{\mcitedefaultendpunct}{\mcitedefaultseppunct}\relax
\EndOfBibitem
\bibitem[Wu and Chai(2015)]{wu:15jctc}
C.-S. Wu and J.-D. Chai, \emph{J. Chem. Theory Comput.}, 2015, \textbf{11},
  2003--2011\relax
\mciteBstWouldAddEndPuncttrue
\mciteSetBstMidEndSepPunct{\mcitedefaultmidpunct}
{\mcitedefaultendpunct}{\mcitedefaultseppunct}\relax
\EndOfBibitem
\bibitem[Smith~Jr(1967)]{smith:67tca}
V.~H. Smith~Jr, \emph{Theor. Chim. Acta (Berlin)}, 1967, \textbf{7}, 245\relax
\mciteBstWouldAddEndPuncttrue
\mciteSetBstMidEndSepPunct{\mcitedefaultmidpunct}
{\mcitedefaultendpunct}{\mcitedefaultseppunct}\relax
\EndOfBibitem
\bibitem[Coulson(1960)]{coulson:60rmp}
C.~A. Coulson, \emph{Rev. Mod. Phys.}, 1960, \textbf{32}, 170\relax
\mciteBstWouldAddEndPuncttrue
\mciteSetBstMidEndSepPunct{\mcitedefaultmidpunct}
{\mcitedefaultendpunct}{\mcitedefaultseppunct}\relax
\EndOfBibitem
\bibitem[Scuseria \emph{et~al.}(2011)Scuseria, Jim{\'e}nez-Hoyos, Henderson,
  Samanta, and Ellis]{scuseria:11jcp}
G.~E. Scuseria, C.~A. Jim{\'e}nez-Hoyos, T.~M. Henderson, K.~Samanta and J.~K.
  Ellis, \emph{J. Chem. Phys.}, 2011, \textbf{135}, 124108\relax
\mciteBstWouldAddEndPuncttrue
\mciteSetBstMidEndSepPunct{\mcitedefaultmidpunct}
{\mcitedefaultendpunct}{\mcitedefaultseppunct}\relax
\EndOfBibitem
\bibitem[Jimenez-Hoyos \emph{et~al.}(2012)Jimenez-Hoyos, Henderson,
  Tsuchimochi, and Scuseria]{jimenez-hoyos:12jcp}
C.~A. Jimenez-Hoyos, T.~M. Henderson, T.~Tsuchimochi and G.~E. Scuseria,
  \emph{J. Chem. Phys.}, 2012, \textbf{136}, 164109\relax
\mciteBstWouldAddEndPuncttrue
\mciteSetBstMidEndSepPunct{\mcitedefaultmidpunct}
{\mcitedefaultendpunct}{\mcitedefaultseppunct}\relax
\EndOfBibitem
\bibitem[Mazziotti(1998)]{mazziotti:98cpl}
D.~A. Mazziotti, \emph{Chem. Phys. Lett.}, 1998, \textbf{289}, 419--427\relax
\mciteBstWouldAddEndPuncttrue
\mciteSetBstMidEndSepPunct{\mcitedefaultmidpunct}
{\mcitedefaultendpunct}{\mcitedefaultseppunct}\relax
\EndOfBibitem
\bibitem[Kutzelnigg and Mukherjee(1999)]{kutzelnigg:99jcp}
W.~Kutzelnigg and D.~Mukherjee, \emph{J. Chem. Phys.}, 1999, \textbf{110},
  2800--2809\relax
\mciteBstWouldAddEndPuncttrue
\mciteSetBstMidEndSepPunct{\mcitedefaultmidpunct}
{\mcitedefaultendpunct}{\mcitedefaultseppunct}\relax
\EndOfBibitem
\bibitem[Pernal and Cioslowski(2004)]{pernal:04jcp}
K.~Pernal and J.~Cioslowski, \emph{J. Chem. Phys.}, 2004, \textbf{120},
  5987--5992\relax
\mciteBstWouldAddEndPuncttrue
\mciteSetBstMidEndSepPunct{\mcitedefaultmidpunct}
{\mcitedefaultendpunct}{\mcitedefaultseppunct}\relax
\EndOfBibitem
\bibitem[L{\"o}wdin and Shull(1956)]{lowdin:56pr}
P.-O. L{\"o}wdin and H.~Shull, \emph{Phys. Rev.}, 1956, \textbf{101},
  1730--1739\relax
\mciteBstWouldAddEndPuncttrue
\mciteSetBstMidEndSepPunct{\mcitedefaultmidpunct}
{\mcitedefaultendpunct}{\mcitedefaultseppunct}\relax
\EndOfBibitem
\bibitem[Bader and Stephens(1974)]{bader:74cpl}
R.~F.~W. Bader and M.~E. Stephens, \emph{Chem. Phys. Lett.}, 1974, \textbf{26},
  445\relax
\mciteBstWouldAddEndPuncttrue
\mciteSetBstMidEndSepPunct{\mcitedefaultmidpunct}
{\mcitedefaultendpunct}{\mcitedefaultseppunct}\relax
\EndOfBibitem
\bibitem[Matito \emph{et~al.}(2007)Matito, Sol\`a, Salvador, and
  Duran]{matito:07fd}
E.~Matito, M.~Sol\`a, P.~Salvador and M.~Duran, \emph{Faraday Discuss.}, 2007,
  \textbf{135}, 325--345\relax
\mciteBstWouldAddEndPuncttrue
\mciteSetBstMidEndSepPunct{\mcitedefaultmidpunct}
{\mcitedefaultendpunct}{\mcitedefaultseppunct}\relax
\EndOfBibitem
\bibitem[Matito \emph{et~al.}(2006)Matito, Duran, and Sol{\`a}]{matito:06jce}
E.~Matito, M.~Duran and M.~Sol{\`a}, \emph{J. Chem. Educ.}, 2006, \textbf{83},
  1243\relax
\mciteBstWouldAddEndPuncttrue
\mciteSetBstMidEndSepPunct{\mcitedefaultmidpunct}
{\mcitedefaultendpunct}{\mcitedefaultseppunct}\relax
\EndOfBibitem
\bibitem[Takatsuka \emph{et~al.}(1978)Takatsuka, Fueno, and
  Yamaguchi]{takatsuka:78tca}
K.~Takatsuka, T.~Fueno and K.~Yamaguchi, \emph{Theor. Chim. Acta (Berlin)},
  1978, \textbf{48}, 175--183\relax
\mciteBstWouldAddEndPuncttrue
\mciteSetBstMidEndSepPunct{\mcitedefaultmidpunct}
{\mcitedefaultendpunct}{\mcitedefaultseppunct}\relax
\EndOfBibitem
\bibitem[Staroverov and Davidson(2000)]{staroverov:00cpl}
V.~N. Staroverov and E.~R. Davidson, \emph{Chem. Phys. Lett.}, 2000,
  \textbf{330}, 161--168\relax
\mciteBstWouldAddEndPuncttrue
\mciteSetBstMidEndSepPunct{\mcitedefaultmidpunct}
{\mcitedefaultendpunct}{\mcitedefaultseppunct}\relax
\EndOfBibitem
\bibitem[Head-Gordon(2003)]{head:03cpl}
M.~Head-Gordon, \emph{Chem. Phys. Lett.}, 2003, \textbf{372}, 508--511\relax
\mciteBstWouldAddEndPuncttrue
\mciteSetBstMidEndSepPunct{\mcitedefaultmidpunct}
{\mcitedefaultendpunct}{\mcitedefaultseppunct}\relax
\EndOfBibitem
\bibitem[Bochicchio \emph{et~al.}(2003)Bochicchio, Torre, and
  Lain]{bochicchio:03cpl}
R.~C. Bochicchio, A.~Torre and L.~Lain, \emph{Chem. Phys. Lett.}, 2003,
  \textbf{380}, 486--487\relax
\mciteBstWouldAddEndPuncttrue
\mciteSetBstMidEndSepPunct{\mcitedefaultmidpunct}
{\mcitedefaultendpunct}{\mcitedefaultseppunct}\relax
\EndOfBibitem
\bibitem[Head-Gordon(2003)]{head:03cpl2}
M.~Head-Gordon, \emph{Chem. Phys. Lett.}, 2003, \textbf{380}, 488--489\relax
\mciteBstWouldAddEndPuncttrue
\mciteSetBstMidEndSepPunct{\mcitedefaultmidpunct}
{\mcitedefaultendpunct}{\mcitedefaultseppunct}\relax
\EndOfBibitem
\bibitem[Mazziotti(2012)]{mazziotti:12cr}
D.~A. Mazziotti, \emph{Chem. Rev.}, 2012, \textbf{112}, 244--262\relax
\mciteBstWouldAddEndPuncttrue
\mciteSetBstMidEndSepPunct{\mcitedefaultmidpunct}
{\mcitedefaultendpunct}{\mcitedefaultseppunct}\relax
\EndOfBibitem
\bibitem[Knowles and Handy(1989)]{knowles:89cpc}
P.~Knowles and N.~Handy, \emph{Comput. Phys. Commun.}, 1989, \textbf{54},
  75\relax
\mciteBstWouldAddEndPuncttrue
\mciteSetBstMidEndSepPunct{\mcitedefaultmidpunct}
{\mcitedefaultendpunct}{\mcitedefaultseppunct}\relax
\EndOfBibitem
\bibitem[Frisch \emph{et~al.}()Frisch, Trucks, Schlegel, Scuseria, Robb,
  Cheeseman, Scalmani, Barone, Mennucci, Petersson, Nakatsuji, Caricato, Li,
  Hratchian, Izmaylov, Bloino, Zheng, Sonnenberg, Hada, Ehara, Toyota, Fukuda,
  Hasegawa, Ishida, Nakajima, Honda, Kitao, Nakai, Vreven, Montgomery, Peralta,
  Ogliaro, Bearpark, Heyd, Brothers, Kudin, Staroverov, Kobayashi, Normand,
  Raghavachari, Rendell, Burant, Iyengar, Tomasi, Cossi, Rega, Millam, Klene,
  Knox, Cross, Bakken, Adamo, Jaramillo, Gomperts, Stratmann, Yazyev, Austin,
  Cammi, Pomelli, Ochterski, Martin, Morokuma, Zakrzewski, Voth, Salvador,
  Dannenberg, Dapprich, Daniels, Farkas, Foresman, Ortiz, Cioslowski, and
  Fox]{g09}
M.~J. Frisch, G.~W. Trucks, H.~B. Schlegel, G.~E. Scuseria, M.~A. Robb, J.~R.
  Cheeseman, G.~Scalmani, V.~Barone, B.~Mennucci, G.~A. Petersson,
  H.~Nakatsuji, M.~Caricato, X.~Li, H.~P. Hratchian, A.~F. Izmaylov, J.~Bloino,
  G.~Zheng, J.~L. Sonnenberg, M.~Hada, M.~Ehara, K.~Toyota, R.~Fukuda,
  J.~Hasegawa, M.~Ishida, T.~Nakajima, Y.~Honda, O.~Kitao, H.~Nakai, T.~Vreven,
  J.~A. Montgomery, {Jr.}, J.~E. Peralta, F.~Ogliaro, M.~Bearpark, J.~J. Heyd,
  E.~Brothers, K.~N. Kudin, V.~N. Staroverov, R.~Kobayashi, J.~Normand,
  K.~Raghavachari, A.~Rendell, J.~C. Burant, S.~S. Iyengar, J.~Tomasi,
  M.~Cossi, N.~Rega, J.~M. Millam, M.~Klene, J.~E. Knox, J.~B. Cross,
  V.~Bakken, C.~Adamo, J.~Jaramillo, R.~Gomperts, R.~E. Stratmann, O.~Yazyev,
  A.~J. Austin, R.~Cammi, C.~Pomelli, J.~W. Ochterski, R.~L. Martin,
  K.~Morokuma, V.~G. Zakrzewski, G.~A. Voth, P.~Salvador, J.~J. Dannenberg,
  S.~Dapprich, A.~D. Daniels, Ö.~Farkas, J.~B. Foresman, J.~V. Ortiz,
  J.~Cioslowski and D.~J. Fox, \emph{Gaussian~09 {R}evision {D}.01}, Gaussian
  Inc. Wallingford CT 2009\relax
\mciteBstWouldAddEndPuncttrue
\mciteSetBstMidEndSepPunct{\mcitedefaultmidpunct}
{\mcitedefaultendpunct}{\mcitedefaultseppunct}\relax
\EndOfBibitem
\bibitem[Hollett and Gill(2011)]{hollett:11jcp}
J.~W. Hollett and P.~M. Gill, \emph{J. Chem. Phys.}, 2011, \textbf{134},
  114111\relax
\mciteBstWouldAddEndPuncttrue
\mciteSetBstMidEndSepPunct{\mcitedefaultmidpunct}
{\mcitedefaultendpunct}{\mcitedefaultseppunct}\relax
\EndOfBibitem
\bibitem[Ramos-Cordoba \emph{et~al.}(2015)Ramos-Cordoba, Lopez, Piris, and
  Matito]{ramos-cordoba:15jcp}
E.~Ramos-Cordoba, X.~Lopez, M.~Piris and E.~Matito, \emph{J. Chem. Phys.},
  2015, \textbf{143}, 164112\relax
\mciteBstWouldAddEndPuncttrue
\mciteSetBstMidEndSepPunct{\mcitedefaultmidpunct}
{\mcitedefaultendpunct}{\mcitedefaultseppunct}\relax
\EndOfBibitem
\bibitem[Fromager(2015)]{fromager:15mp}
E.~Fromager, \emph{Molec. Phys.}, 2015, \textbf{113}, 419--434\relax
\mciteBstWouldAddEndPuncttrue
\mciteSetBstMidEndSepPunct{\mcitedefaultmidpunct}
{\mcitedefaultendpunct}{\mcitedefaultseppunct}\relax
\EndOfBibitem
\end{mcitethebibliography}
\end{document}